\documentclass[%
 reprint,
 amsmath,amssymb,
 aps,
]{revtex4-2}

\usepackage{natbib}
\usepackage{multirow}

\usepackage{graphicx}% Include figure files
\usepackage{dcolumn}% Align table columns on decimal point
\usepackage{bm}% bold math
\usepackage{hyperref}% add hypertext capabilities

\usepackage{booktabs}
\usepackage{threeparttable}
\usepackage{color}
\usepackage{xcolor}

\begin{document}

\title{Probing Dark Matter Substructure with Image Number Anomaly in Strong Lensing Systems}
\author{Wenlin Hou$^{1}$}
\author{Jianxiang Liu$^{2, 3}$}
\author{Kai Liao$^{1}$}%
\email{E-mail:liaokai@whu.edu.cn}
\affiliation{$^{1}$School of Physics and Technology, Wuhan University, Wuhan 430072, China\\
$^{2}$National Astronomical Observatories, Chinese Academy of Sciences, Beijing 100101, P. R. China\\
$^{3}$School of Astronomy and Space Sciences, University of Chinese Academy of Sciences, Beijing 100049, P. R. China
}

\date{\today}% It is always \today, today,
             %  but any date may be explicitly specified

\begin{abstract}

Gravitational lensing observables, including anomalies in image positions, flux ratios, and time delays, serve as usual probes of dark matter (DM) substructure. When dark matter substructure possesses sufficient perturbations, it may lead to the formation of extra images in otherwise canonical doubly or quadruply imaged systems. With the advent of increasingly precise observational instruments, previously undetectable images may become measurable and image number anomalies therefore could  be an increasingly viable method. In this paper, we utilize the gravitational lensing phenomenon of image number anomaly to derive constraints on dark matter substructure. We present the extra images induced by distinct forms of DM substructure, specifically primordial black holes (PBHs) and fuzzy dark matter (FDM) and show that higher angular resolution observations increase the probability of detecting additional lensed images. Based on a null detection of image number anomalies in a sample of 3500 lens systems generated from the \textit{Strong Lensing Halo model-based mock catalogs} (SL-Hammocks), we derive upper limits on the abundance of PBHs. At the 95\% confidence level, the PBH abundance is constrained to $\lesssim 0.125\%$, $0.08\%$, and $0.04\%$ for PBH masses in the range $\sim 10^{7}$--$10^{9}~M_{\odot}$, corresponding to angular resolutions of $0.1''$, $0.05''$, and $0.01''$, respectively. Similarly, we exclude particle masses below $0.4$, $0.6$, and $3.5 \times 10^{-22} \ \mathrm{eV}$ for FDM at the same confidence level for the respective resolutions. Furthermore, the abundance of PBHs $\lesssim 0.9\%$ could be constrained at an angular resolution of $0.5''$ for the Legacy Survey of Space and Time (LSST) Observations. Finally, we discuss methodologies for identifying image number anomalies in special cases and demonstrate feasibility using a fitting procedure.

\end{abstract}

\maketitle

\section{Introduction}

Dark matter, a hypothetical form of matter, is estimated to be approximately five times more abundant than visible matter in the universe. Its existence is supported by multiple lines of evidence, including galactic rotation curves \cite{Rubin1978} and gravitational lensing by galaxy clusters \cite{Clowe2006}. While the precise nature of dark matter remains unknown, the Cold Dark Matter (CDM) paradigm has proven successful in explaining the formation of large-scale structures in the universe \cite{Blumenthal1984}. However, at small scales, the CDM model encounters non-trivial challenges. Notably, it fails to account for observational phenomena such as the missing satellites problem \cite{Klypin1999,Moore1999} and the cusp--core discrepancy \cite{deBlok2010}, which refer respectively to the overprediction of dwarf galaxy abundances and the preference of cored, rather than cuspy, dark matter halo profiles in observations. To investigate the fundamental nature and constituent properties of dark matter and address the persistent discrepancies, numerous alternative dark matter candidates, which remain the subject of active theoretical and experimental research efforts, have been proposed. Among them, primordial nlack holes (PBHs) \cite{Carr2020} and fuzzy dark matter (FDM) \cite{Hu2000,Hui2017} have attracted considerable interest. PBHs are hypothesized to have formed in the early universe via a variety of mechanisms, and their broad mass spectrum, ranging from sub-planetary scales to supermassive scales, makes them a particularly versatile candidate. PBHs as lenses within foreground galaxy clusters have been probed using caustic crossing method by observing highly magnified stars in the background galaxies\cite{2018NatAs...2..334K,2018PhRvD..97b3518O,2023A&A...679A..31D}. In these works, lens masses smaller than $10^6M_\odot$ were considered while we focus on millilensing with substructure mass scale $>10^6M_\odot$ in galaxy lenses. In contrast, FDM, characterized by its ultra-light mass and wave-like behavior, can naturally suppress small-scale structure formation and prevent the development of overly dense galactic cores, offering a promising resolution to the small-scale challenges faced by CDM. 
Recent lensing studies show FDM (also called wave dark matter) accurately predicts the brightness and position anomalies of the multiple lensed images of the HS 0810+2554 lens system \cite{2023NatAs...7..736A}.

A central challenge in the study of dark matter lies in understanding the small-scale structure problem, wherein the standard CDM model struggles to account for observed features on sub-galactic scales \cite{Bullock2017}. Strong gravitational lensing offers a powerful tool to investigate such small-scale structures across cosmological distances \cite{Vegetti2009}. While the CDM subhalos perturb the critical lines and can be probed with the anomalous position of highly magnified stars in cluster lensing\cite{2018ApJ...867...24D,2024ApJ...961..200W}, for example $10^7-10^9M_\odot/h$ subhalos or FDM with mass $\sim10^{-24}$ could explain the event “Mothra” \cite{2024PhRvD.109h3517A}, a widely adopted technique focuses on analyzing flux ratio anomalies in quadruply lensed quasar systems \cite{Mao1998,2010AdAst2010E...9Z}, which are sensitive to perturbations from low-mass subhalos. While this method benefits from the relatively large number of known quadruply lensed systems, it is often hindered by stellar microlensing, which can mimic or obscure substructure signals \cite{Schechter2002}. To mitigate these complications, observations are frequently conducted at radio, mid-infrared wavelengths or [OIII] narrow-line band originated from larger regions, where stellar microlensing effects are significantly reduced \cite{2009ApJ...699.1578M, 2015MNRAS.454..287J}. However, agreement on the nature of dark matter has not yet been reached.
Besides, dark matter substructures can be probed through the detection of subtle anomalies time delays \cite{Keeton2009, Liu2024} and positions \cite{2010AdAst2010E...9Z}. These signatures are becoming increasingly accessible as observational capabilities continue to improve. 

In addition to these well-studied effects, substructures acting as secondary lenses may induce additional images, which have remained largely unexplored due to the limited spatial resolution of earlier instruments, other than a paper mention that DM substructure would increase fraction of quads in surveys \cite{2004ApJ...608...25C}. However, recent advancements in observational precision now allow such effects to be investigated. For example, the James Webb Space Telescope (JWST) offers spatial resolutions of approximately 0.03 to 0.06 arcseconds in the near-infrared \cite{Rigby2022}, while radio observations using Very Long Baseline Interferometry (VLBI) can achieve sub-milliarcsecond resolution, depending on the configuration \cite{Reid2014}. Furthermore, transient sources such as Fast Radio Bursts (FRBs) and Gravitational Waves (GWs) can now be resolved with high temporal precision \cite{Liao2017,2019RPPh...82l6901O,Liao:2022}, opening new avenues for time-domain lensing studies. With these enhanced observational tools, it is increasingly feasible to detect extra images caused by dark matter substructure. Our research aims to leverage these capabilities to investigate the small-scale structure of dark matter through the analysis of image number anomalies in strong gravitational lensing systems.

This paper is structured as follows. Section\ref{Sec II} shows the potential occurrence of image number anomalies in lensing systems subject to primordial black hole perturbations, as well as the impact of fuzzy dark matter. Section \ref{Sec III} presents the generation of strong lensing systems and derives constraints on two distinct classes of dark matter substructure. In Section \ref{Sec IV}, we analyze the identification of image number anomalies in special cases. A summary and discussion are provided in Section \ref{Sec V}. Through out the paper we use the best fit results \cite{2020A&A...641A...6P} obtained by Planck in 2018 ($H_0=67.7\mathrm{\,km\,s^{-1}\,Mpc^{-1}}$, $\Omega_\mathrm{m}=0.31$).

\section{Multiple images by substructures} \label{Sec II}

Our model intentionally neglects the contribution of line-of-sight structures and FDM subhalos since they are typically insufficient to induce the magnitude of image number anomalies and focuses on two representative types of dark matter substructure: primordial black holes and FDM-induced density fluctuations.

For clarity and simplicity, we present the image number anomalies due to perturbations from PBHs and FDM, modeled within a smooth macrolens framework that separately accounts for baryonic and dark matter components. The dark matter halo is represented by an elliptical Navarro–Frenk–White (NFW) profile \cite{1996ApJ...462..563N, 2021PASP..133g4504O}, a standard choice in gravitational lensing based on N-body simulations. The baryonic component is modeled with an elliptical Hernquist profile \cite{1990ApJ...356..359H}. This two-component model effectively isolates the dark and baryonic mass contributions. External shear is taken into consideration. All simulations are conducted using the \texttt{lenstronomy} Python package \cite{2018ascl.soft04012B, 2021JOSS....6.3283B}, which supports a broad range of lens and source models.

\begin{table*}[htbp]
\centering
\caption{Lens model parameters (lenstronomy notation and units).}
\label{tab:lens_params_correct_units}
\begin{tabular}{llll}
\toprule
Component & Parameter & Symbol & Value (units) \\
\midrule
\multirow{5}{*}{\textbf{Main Halo (Elliptical NFW)}} 
  & Scale radius & $R_{\rm s1}$ & $2.77\ \mathrm{arcsec}$ \\
  & Deflection at $R_{\rm s1}$ & $\alpha_{R_{\rm s1}}$ & $0.37\ \mathrm{arcsec}$ \\
  & Ellipticity & $e_1$ & $-0.14$ \\
  &               & $e_2$ & $-0.12$ \\
  & Center & $(x,y)$ & $(0,\ 0)\ \mathrm{arcsec}$ \\
\midrule
\multirow{5}{*}{\textbf{Baryonic / Stellar (Elliptical Hernquist)}} 
  & Surface-density norm & $\sigma_0$ & $9.9\ \ (\Sigma/\Sigma_{\rm crit},\ \text{dimensionless})$ \\
  & Scale radius & $R_{\rm s2}$ & $0.09\ \mathrm{arcsec}$ \\
  & Ellipticity & $e_1$ & $-0.024$ \\
  &               & $e_2$ & $-0.026$ \\
  & Center & $(x,y)$ & $(0,\ 0)\ \mathrm{arcsec}$ \\
\midrule
\multirow{3}{*}{\textbf{External Shear}}
  & Shear amplitude & $\gamma_{\rm ext}$ & $0.08$ \\
  & Shear angle & $\psi_{\rm ext}$ & $1.45\ \mathrm{rad}$ \\
  & Reference center & $(x,y)$ & $(0,\ 0)\ \mathrm{arcsec}$ \\
\bottomrule
\end{tabular}

\vspace{1ex}
\footnotesize
\emph{Notes:} In lenstronomy the Hernquist ``$\sigma_0$'' used in the lens model is the dimensionless scaled surface density,
$\sigma_0 = \Sigma_{\rm phys}/\Sigma_{\rm crit}$.  \\ \label{macro}
\end{table*}

The NFW profile, commonly employed to model the density
distribution of dark matter halos, is given by
\begin{equation}
    \rho_{\mathrm{NFW}}(r) =
    \frac{\rho_{\rm s}}{(r/r_{\rm s})\,(1+r/r_{\rm s})^{2}},
\end{equation}
where $ r $ denotes the three-dimensional radial distance, $ r_{\rm s} $ is
the physical scale radius, and $ \rho_{\rm s} $ is the characteristic density. The luminous component follows the Hernquist profile whose three-dimensional density distribution is
\begin{equation}
    \rho_{\mathrm{H}}(r) = \frac{M_{\mathrm{H}}}{2\pi}
    \frac{r_{\rm s}}{r\,(r+r_{\rm s})^{3}},
\end{equation}
where $M_{\mathrm{H}}$ is the total stellar mass and $r_{\rm s}$ is the
three-dimensional scale radius. As with the NFW component, we adopt an
elliptical projected potential to capture the intrinsic shape of the stellar
distribution. In accordance with the conventions adopted in \texttt{lenstronomy}, we describe
the projected Hernquist component using the parameters $\sigma_{0}$ and $R_{\rm s2}$ rather than the physical parameters $M_{\mathrm{H}}$ and $r_{\rm s}$. The Einstein radius, $\theta_{\mathrm{E}}$, is defined as the angular radius within which the mean surface mass density of the lens equals the critical surface mass density for lensing,
\begin{equation}
\bar{\Sigma}(<\theta_{\mathrm{E}}) = \Sigma_{\mathrm{crit}},
\end{equation}
where
\begin{equation}
\Sigma_{\mathrm{crit}} = \frac{c^{2}}{4\pi G} \frac{D_{\mathrm{s}}}{D_{\mathrm{d}} D_{\mathrm{ds}}},
\end{equation}
with $D_{\mathrm{d}}$, $D_{\mathrm{s}}$, and $D_{\mathrm{ds}}$ denoting the angular diameter distances to the lens, to the source, and between the lens and source, respectively. The source position is fixed at $(x_{\mathrm{s}}, y_{\mathrm{s}}) = (0.04, -0.03)$ in the source plane. The redshifts of the source and lens are held constant at $z_{\mathrm{source}} = 1.74$ and $z_{\mathrm{lens}} = 0.789$, respectively. The remaining principal parameters characterizing the smooth macrolens model are provided in Table \ref{macro}.

To identify the emergence of additional lensed images, we impose two criteria. First, the total number of images must exceed four for quad systems or two for double systems. Second, the angular separation between every pair of images must be larger than the adopted spatial resolution threshold. 
The spatial resolution of the JWST ranges from approximately $0.03''$ to $0.06''$, 
which lies within a broader representative interval of [$0.01''$, $0.1''$]. 
For quantitative analysis, we consider three characteristic resolutions: $0.1''$, $0.05''$, and $0.01''$.

\subsection{Supermassive PBH perturbations}

To incorporate PBHs into the smooth lens model, we posit that a fraction of the dark matter consists of PBHs, modeled as point masses. The PBH mass function is assumed to follow a power law, $dN/dm \propto m^{-1.8}$, spanning a supermassive range from $10^{7} M_\odot$ to $10^{9} M_\odot$. The spatial distribution of these PBHs is modeled to trace the overall NFW dark matter halo, such that their convergence is given by $\kappa_{\mathrm{PBH}}(r) = f_{\mathrm{PBH}} \cdot \kappa_{\mathrm{NFW}}(r)$, with the PBH fraction fixed at $f_{\mathrm{PBH}} = 0.005$. The PBH mass, $M_{\mathrm{PBH}}$, is represented by the median of each considered mass interval. We set it equal to $10^9 M_{\odot}$ for the interval $[10^8 M_{\odot}, 10^{10} M_{\odot}]$). Because multiple images typically form near the critical curves, which are characterized by the Einstein radius $\theta_\mathrm{E}$, we randomly sample positions of individual PBHs within a projected radius of $2\theta_\mathrm{E}$ from the lens center. PBHs located within this area are those most likely to influence the formation of multiple images. A compensating negative-mass sheet is introduced to enforce global mass conservation within $2\theta_\mathrm{E}$.

\begin{figure*}
 \includegraphics[width=5cm,angle=0]{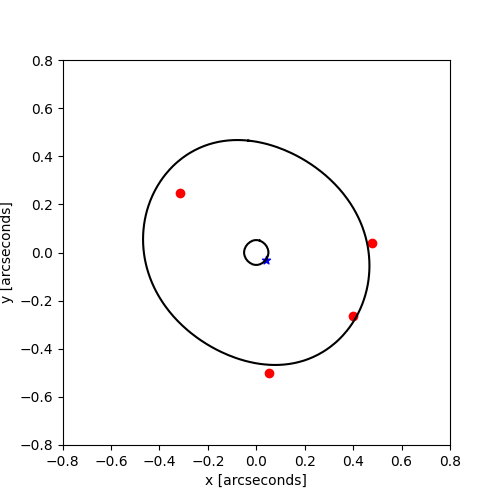}
 \includegraphics[width=5cm,angle=0]{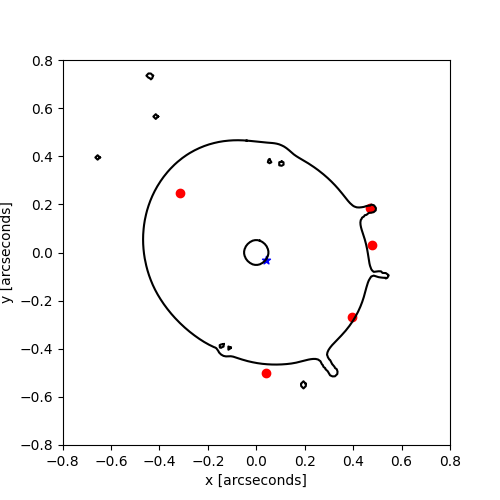}
 \includegraphics[width=5cm,angle=0]{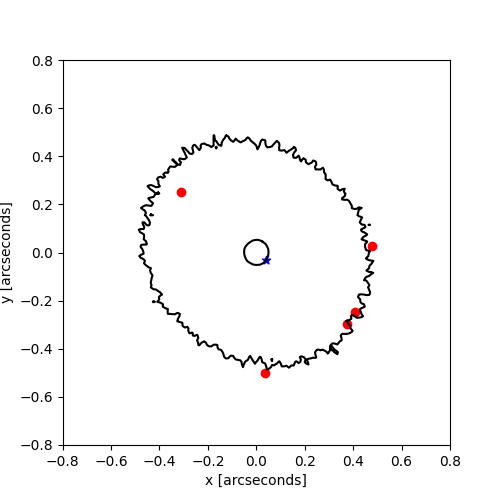}
  \caption{Lensed image numbers and critical curves for a strong lensing system with parameters in Table \ref{macro}. The left panel shows original quad system, i.e., without dark matter substructure. The blue star denotes the fixed source position. The red points indicate the lensed images. The black solid curves correspond to the critical curves. The middle panel presents the image number anomaly induced by PBHs. The right panel shows the one arising from FDM-induced fluctuations corresponding to an ultra-light boson mass of $m_{\psi} = 10^{-22}\,{\mathrm{eV}}$. All spatial units are given in arcseconds.
  } \label{pbhfdm}
\end{figure*}

As left panel of Fig. \ref{pbhfdm} shows, a cusp quadruple-image configuration is obtained from the smooth macro-model. The resulting 5 lensed images which induced by PBHs for angular resolution $0.01''$, depicted in the middle panel of Fig. \ref{pbhfdm}, demonstrate that the inclusion of PBH substructure perturbs the critical curves, thereby inducing measurable anomalies in the image numbers. 

\begin{figure}
	\centering
	\begin{minipage}{\linewidth}
		\centering
		\includegraphics[width=\linewidth]{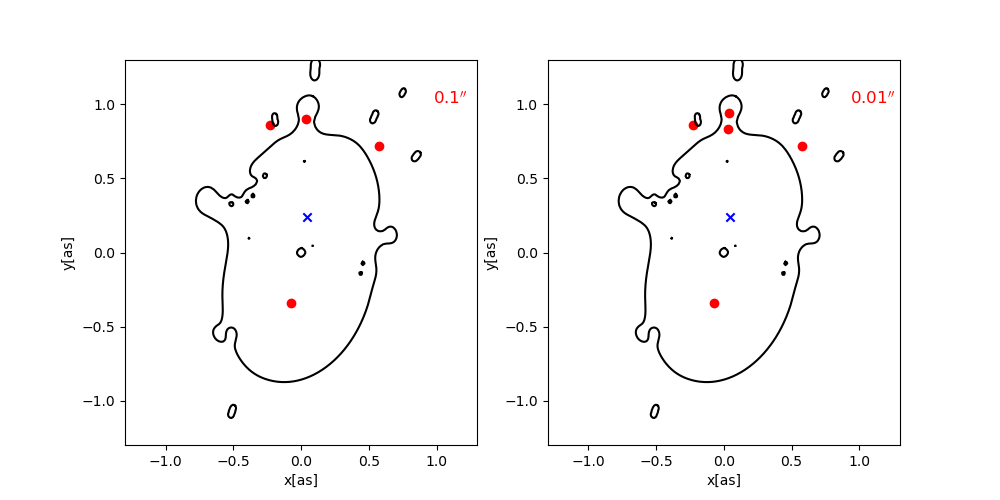}	
	\end{minipage}
    \caption{Effect of angular resolution on the number of observed lensed images. The left panel (angular resolution $0.1''$) and the right panel (angular resolution $0.01''$) show a clear increase in the number of resolved images with higher resolution, observing 4 and 5 images, respectively. Observed lensed images are marked by red dots, while the true source position is indicated by a blue cross. The black solid lines denote critical curves for the lens model.
    }
    \label{fig:resolution}
\end{figure}

Fig. \ref{fig:resolution} illustrates the dependence of the number of gravitationally lensed images on the angular resolution of the observation. Higher angular resolution enables the detection of a greater number of images. In the right panel of Fig.~\ref{fig:resolution}, where the angular resolution is set to $0.01''$, images separated by angular distances greater than $0.01''$ can be resolved. In contrast, the left panel, with an angular resolution of $0.1''$, shows that images separated by less than $0.1''$ are merged together. For candidates separated by less than $0.1''$ we replaced the individual positions by a single, flux‑weighted centroid for illustration. This differs from default behavior of \texttt{lenstronomy}, which discards redundant candidates that lie within the image distance threshold.

For the showcase lens system, multiple sets of 100 simulations are performed at each angular resolution. The mean value across these simulations is adopted to quantify the frequency of image-number anomalies within a batch of 100 runs. The results are presented in left column of Table \ref{0.005}. We note that a magnification threshold of 0.1 is applied to exclude the centrally demagnified image. As presented in left column of Table \ref{0.005}, the incidence of image-number anomalies increases with improving spatial resolution. which is consistent with the expectation that higher resolutions enable the detection of images that remain unresolved at lower resolutions. An angular resolution of $0.005''$ is adopted as the practical limit for image distance measurements in consideration of computation complexity and runtime feasibility.

% \begin{table*}
\begin{table}
\renewcommand{\arraystretch}{1.5}
\begin{tabular}{cccc}
  \hline\hline
  Image distance  &\quad Count (PBHs) &\quad Count (FDM)  \\  % Table header
  \hline
  \hline
  $>0.1''$   & 0.7 & 0 \\  %第一行数据
  \hline
  $>0.05''$   & 1.6 & 0 \\  %第一行数据
  \hline
  $>0.01''$  & 5.3 & 5.8 \\
  \hline
  $>0.005''$  & 5.7 & 6.5 \\
  \hline\hline
\end{tabular}
 \caption{Counts of image number anomalies observed across 100 simulations, stratified by different angular resolutions (image distance). Two distinct columns are included, corresponding to simulations where dark matter substructures are modeled as PBHs and FDM, respectively.
}\label{0.005}
\end{table}

\subsection{FDM-induced fluctuations}

Within the framework of the same macromodel, we examine dark matter substructure composed of fuzzy dark matter. The FDM paradigm posits that dark matter consists of ultralight bosons (with mass $m_{\psi} \sim 10^{-22}$ eV), whose macroscopic de Broglie wavelength---extending to kiloparsec scales---permits a description as a classical scalar field \cite{Hui2017}. A key consequence of this wave description is the emergence of stochastic density fluctuations within the halo on the scale of the de Broglie wavelength $\lambda_{\mathrm{dB}}$.
 
The characteristic scale of these fluctuations, $\lambda_{\mathrm{dB}}$, is set by the boson mass $m_\psi$ and the halo mass $M_h$ according to the relation \cite{2014PhRvL.113z1302S,2016ApJ...818...89S}:
\begin{equation}
\label{equation:lambda}
\lambda_{\mathrm{dB}} = 150 \left( \frac{10^{-22}\,\mathrm{eV}}{m_\psi} \right) \left( \frac{M_h}{10^{12} M_\odot} \right)^{-1/3} \,\mathrm{pc}.
\end{equation}
These fluctuations result from wave interference, causing the local three-dimensional density to vary between zero (destructive interference) and approximately twice the local mean density (constructive interference).
 
The variance of the surface mass density, $\sigma_\Sigma^2$, at a projected position $\boldsymbol{\xi}$ is derived by integrating the three-dimensional density variance along the line of sight. This is given by \cite{2023NatAs...7..736A,2020PhRvL.125k1102C}:
\begin{equation}
\label{equation:sigma_sim}
\sigma_\Sigma^2(\boldsymbol{\xi}) = \lambda_{\mathrm{dB}} \sqrt{\pi} \int_{-\infty}^{\infty} \rho_{\mathrm{smooth}}^2(z,\boldsymbol{\xi})\, \mathrm{d}z.
\end{equation}
Here, $z$ is the line-of-sight coordinate, $\boldsymbol{\xi}$ is the two-dimensional position vector in the plane of the sky, and $\sigma_\rho$ is the standard deviation of the three-dimensional density fluctuations at point $(z,\boldsymbol{\xi})$. This expression directly links the variance of the projected substructure to the fundamental properties of the FDM particle ($m_\psi$, via $\lambda_{\mathrm{dB}}$) and the global halo structure ($\rho_{\mathrm{smooth}}$).

\begin{figure}
\includegraphics[scale=0.5]
{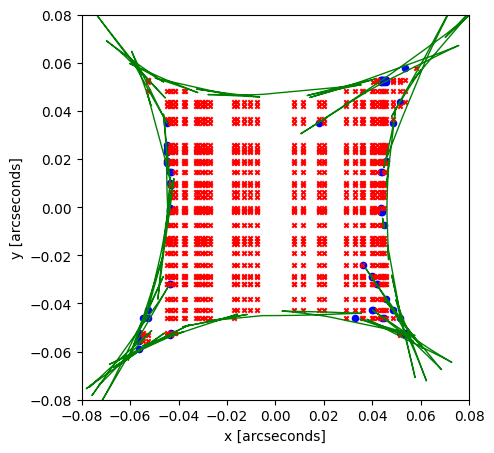}
\caption{Source positions leading to image number anomalies related to FDM-induced perturbations. The caustics, as perturbed by the FDM, are illustrated by the green line. The red crosses denote source positions corresponding to 4 images, while the blue points indicate source positions yielding more than 4 images, i.e., image number anomalies, all within the identical FDM simulation.}\label{fdmedge}
\end{figure}

The surface mass density contribution from FDM substructure is modeled by generating a modulated white noise field, as defined by the variance in Equation (\ref{equation:sigma_sim}), and convolving it with a two-dimensional Gaussian kernel. The kernel's full width at half maximum is set by the de Broglie wavelength, $\lambda _{\mathrm{dB}}$. The total projected mass distribution is the sum of this FDM component and a smooth macro-model. For this simulation, the de Broglie wavelength is set to $\lambda_{\mathrm{dB}} = 150\ \mathrm{pc}$, corresponding to an axion mass of $m_{\psi} = 10^{-22}\ \mathrm{eV}$.

As illustrated in the right panel of Fig. \ref{pbhfdm}, the perturbations induced by FDM clearly lead to fluctuations in the critical lines and generate an image-number anomaly based on the identical cusp quadruple-image configuration. Notably, the resultant critical line morphology distinctly differs from that observed under the influence of PBH perturbations.

For the FDM, we likewise conducted multiple sets of 100 simulations and computed the average number of image-number anomalies observed. As presented in the right column of Table \ref{0.005}, the results also indicate an increase in the frequency of image-number anomalies with decreasing image distance. It is noted that the FDM simulations at resolutions of $0.1''$ yield zero image number anomaly and FDM perturbation basically lose effectiveness at resolution $0.5''$. This is a consequence of the relatively weak perturbation introduced by the FDM in this particular lens system.

We further highlight that the occurrence of multiple-image formation in the FDM is highly sensitive to the source position. Sources located near caustics are particularly prone to generating extra images, owing to the substantially amplified perturbations experienced at these positions, as demonstrated in Fig. \ref{fdmedge}. Furthermore, double-lens systems rarely produce image number anomalies. This behavior aligns with the findings reported in Ref.~\cite{2020PhRvL.125k1102C}. Note that  Ref.~\cite{2020PhRvL.125k1102C} also showed FDM lenses could generate multiple images. In this work, we will demonstrate that the image number anomaly can be used to constrain the properites of dark matter substructure.

\section{constraining substructures with observations} \label{Sec III}

\begin{figure}	\centering	\begin{minipage}{\linewidth}		\centering		\includegraphics[width=\linewidth]{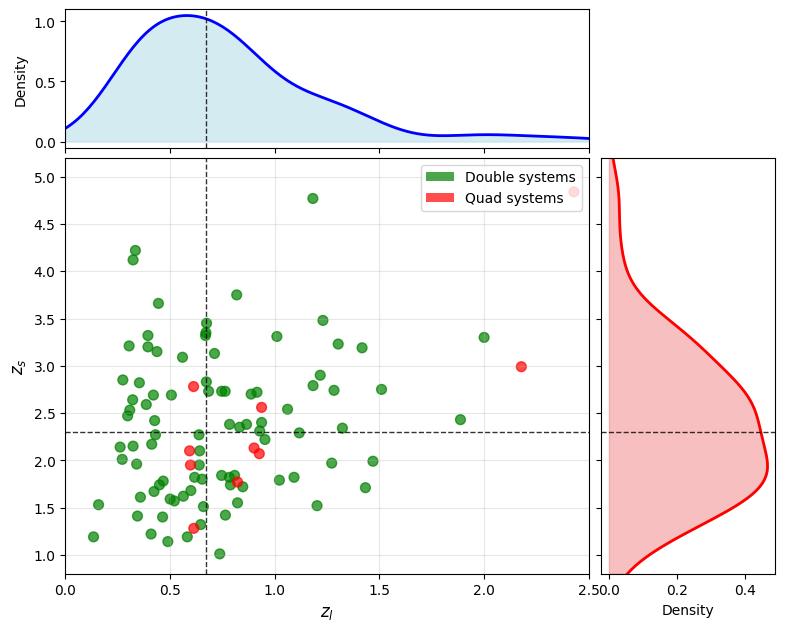}		\end{minipage}    \caption{Redshift distribution of all strong gravitational lensing systems in the 100 samples. Systems classified as doubles and quads are denoted by green and red points, respectively. The median lens and source redshifts are indicated by black dashed lines. Marginal panels (upper and right) present the probability density functions of the lens and source redshift distributions, derived using Gaussian kernel density estimation.}    \label{zlzs}\end{figure}

\begin{figure}	\centering	\begin{minipage}{\linewidth}		\centering		\includegraphics[width=\linewidth]{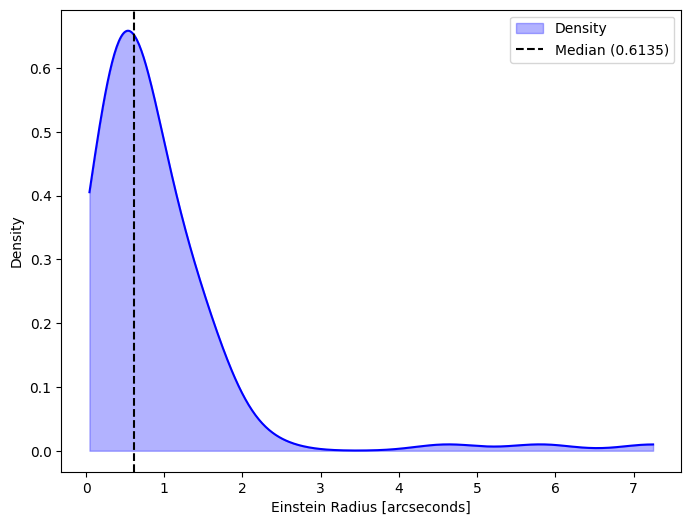}		\end{minipage}    \caption{Probability density curve of the Einstein radii, obtained by applying a Gaussian kernel density estimator to the selected 100 strong lensing systems. The black dashed line indicates the median of the Einstein radii.}    \label{ein_r}\end{figure}

To get observational data for strong gravitational lens systems, we sample 100 lens systems from the LSST mock catalog of gravitationally lensed quasars generated in \cite{2025OJAp....8E...8A}, comprising 90 double-image systems and 10 quadruple-image systems. The SL-Hammocks decouples the dark matter and baryonic components and facilitates the incorporation of fluctuations induced by PBHs and FDM. Fig. \ref{zlzs} and Fig. \ref{ein_r} display the redshift distribution and Einstein radius distribution of the 100 lens systems, respectively.  Image-number anomalies induced by FDM arise only when source positions lie near the edges of caustics, as illustrated in Fig. \ref{fdmedge}. To ensure sufficient coverage of such critical source positions, we perform random sampling of source positions repeatedly. This process yields 35 distinct source position configurations, resulting in a total of 3500 strong lensing systems, including 3150 double-image systems and 350 quadruple-image systems.

From Poisson statistics, the probability of observing $ k $ events given an expected mean rate $ \lambda $ is 
\begin{equation}
P(k \,|\, \lambda) = \frac{\lambda^k e^{-\lambda}}{k!}.
\end{equation}
In the case of a null detection ($ k = 0 $), for a 95\% confidence level $ P(0| \lambda_{\text{up}}) = 0.05 $ and $ \lambda_{\mathrm{up}} \simeq 3.0 $. Given that no lens system with anomalous extra images were detected in the observations, we can exclude, at the 95\% confidence level, the regions of the $(M_{\mathrm{PBH}}, f_{\mathrm{PBH}})$ parameter space that predict at least 3 lens systems with image-number anomalies. To ensure statistical robustness, the Poisson parameter is estimated as the mean value derived from an ensemble of simulations.

\begin{figure}	\centering	\begin{minipage}{\linewidth}		\centering		\includegraphics[width=\linewidth]{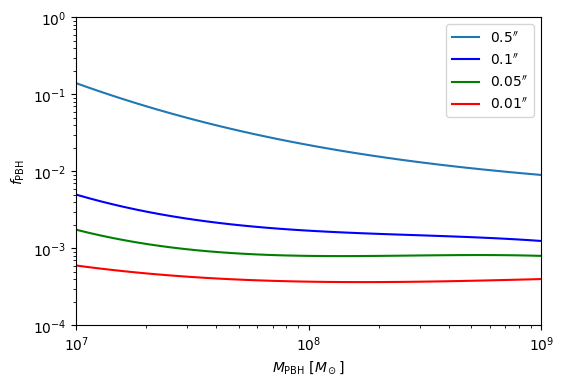}		\end{minipage}    \caption{Upper limits (95\% confidence level) on the abundance of PBHs as a function of mass, derived from the absence of anomalous image number counts in a sample of 3500 lensed systems. Constraints are shown for three angular resolution thresholds: $0.01''$, $0.05''$, and $0.1''$.
}    \label{fig:pbh_constraints}\end{figure}

As illustrated in Fig.~\ref{fig:pbh_constraints}, based on a null detection of image number anomalies in a sample of 3500 lens systems, we derive upper limits on the abundance of PBHs. The PBH abundance is constrained to $ \lesssim 0.125\% $, $ 0.08\% $, and $ 0.04\% $ for PBH masses in the range $ \sim 10^{7} $--$ 10^{9}~M_{\odot} $ at the 95\% confidence level, corresponding to angular resolutions of $ 0.1'' $, $ 0.05'' $, and $ 0.01'' $, respectively. The PBH mass, $M_{\mathrm{PBH}}$, is represented by the median of each considered mass interval (e.g., $10^8 M_{\odot}$ for the interval $[10^7 M_{\odot}, 10^9 M_{\odot}]$). Within the mock catalog, a subset of strong lensing systems is characterized by large Einstein radii. The numerical solution of the lens equations becomes computationally prohibitive and time-consuming in the limit of small PBH masses for such systems. Consequently, the present analysis is restricted to the PBH mass range of $[10^7 M_{\odot}, 10^9 M_{\odot}]$ since the aim of this work is to propose a method of image number anomalies.

We note that the upper limits is determined by several competing factors. On the one hand, $f_{\mathrm{PBH}}$ generally decreases for more massive PBHs, which cause much more pertubations and are more likely to produce detectable image splitting. On the other hand, the complete absence of PBHs with masses around $10^9\,M_\odot$ can occur in relatively small Einstein radii lens system for small $f_{\mathrm{PBH}}$. In such cases, only less massive PBHs (e.g., around $10^8\,M_\odot$) would be present. At this lower mass scale, a greater number of lens systems can be perturbed sufficiently to produce extra images, allowing the PBH abundance fraction $f_{\rm PBH}$ to be constrained, and potentially reduced, even further.

\begin{figure}	\centering	\begin{minipage}{\linewidth}		\centering		\includegraphics[width=\linewidth]{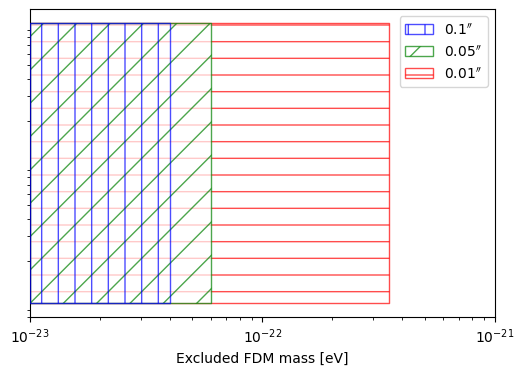}		\end{minipage}    \caption{FDM constraints for 3500 lens systems. Red horizontal lines, green slash lines and blue vertical lines represent spatial resolution $0.01'', 0.05'', 0.1''$ respectively at $95\%$ CL.}    \label{fig:fdm_constraints}\end{figure}

Similarly, constraints on the FDM mass $m_{\psi}$ can be derived, as shown in Fig.~\ref{fig:fdm_constraints}. we exclude particle masses below $ 0.4 $, $ 0.6 $, and $ 3.5 \times 10^{-22} \ \mathrm{eV} $ at the same confidence level for the respective resolutions. Higher spatial resolution enables the detection of more image-number anomalies, therefore more excluded regions.

For LSST observations \cite{2019ApJ...873..111I, 2025OJAp....8E...8A}, the abundance of PBHs can be constrained to $\lesssim 0.9\%$ in the mass range $M \geq 10^{9}~M_{\odot}$ at the 95\% confidence level, assuming an angular resolution of $0.5''$, as illustrated in Fig. \ref{fig:pbh_constraints}. We note that gravitational lensing systems with large Einstein radii exert a more pronounced influence at low spatial resolution and the dominant factor is PBH mass.

\section{Identifying anomalies in some special cases} \label{Sec IV}

\begin{figure*}
\includegraphics[scale=0.35]
{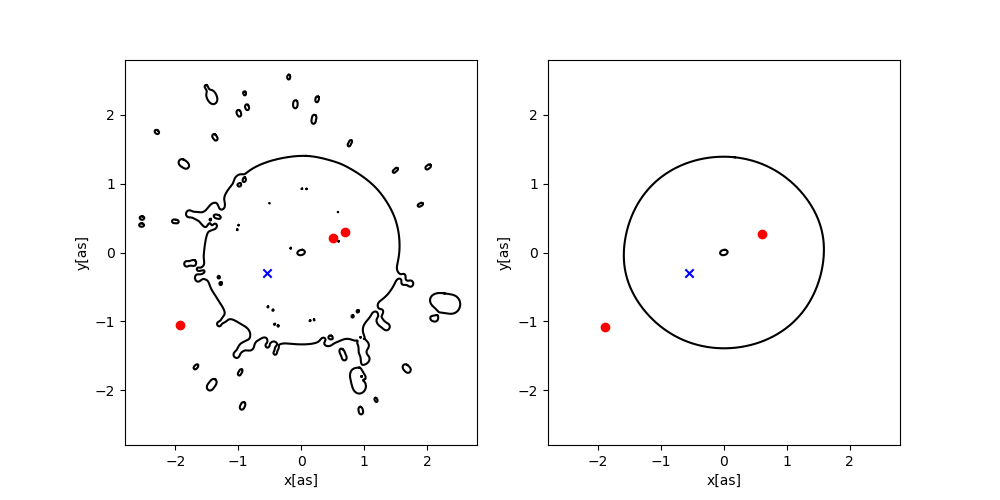}
\includegraphics[scale=0.35]{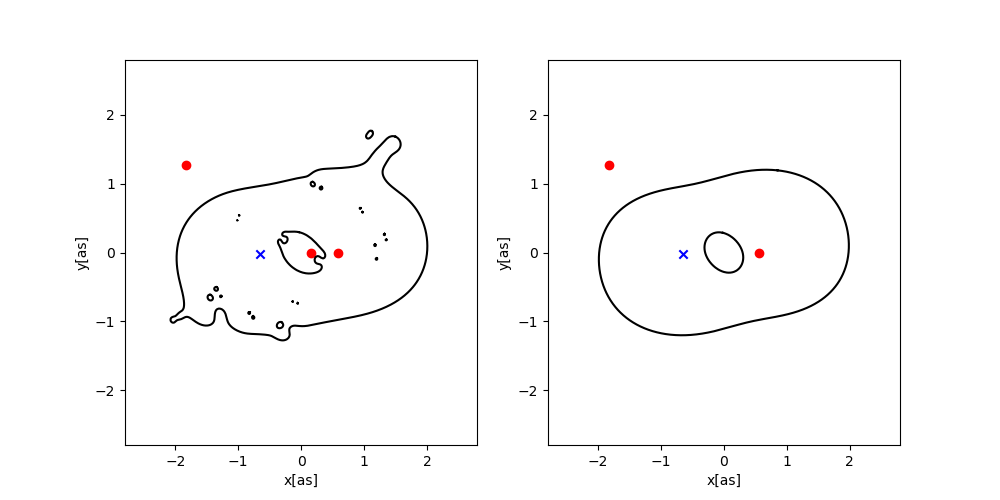}

\includegraphics[scale=0.35]{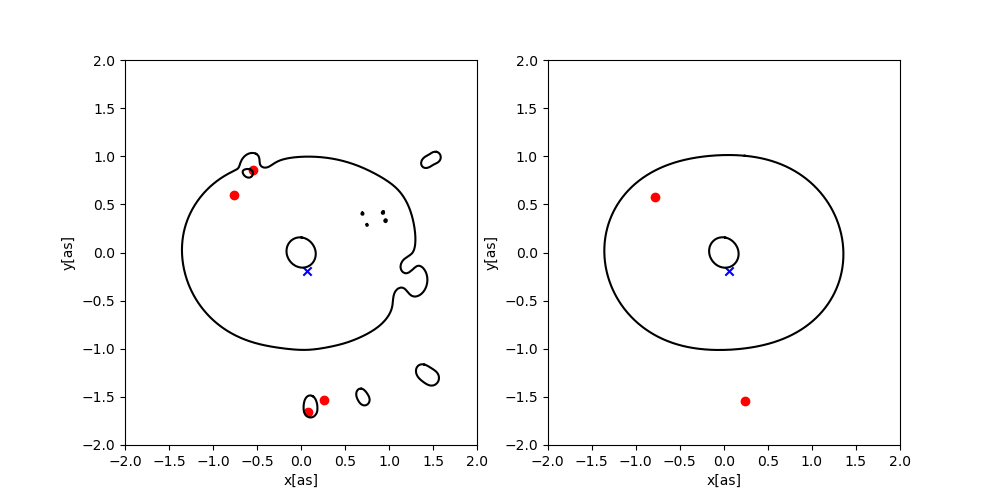}
\includegraphics[scale=0.35]{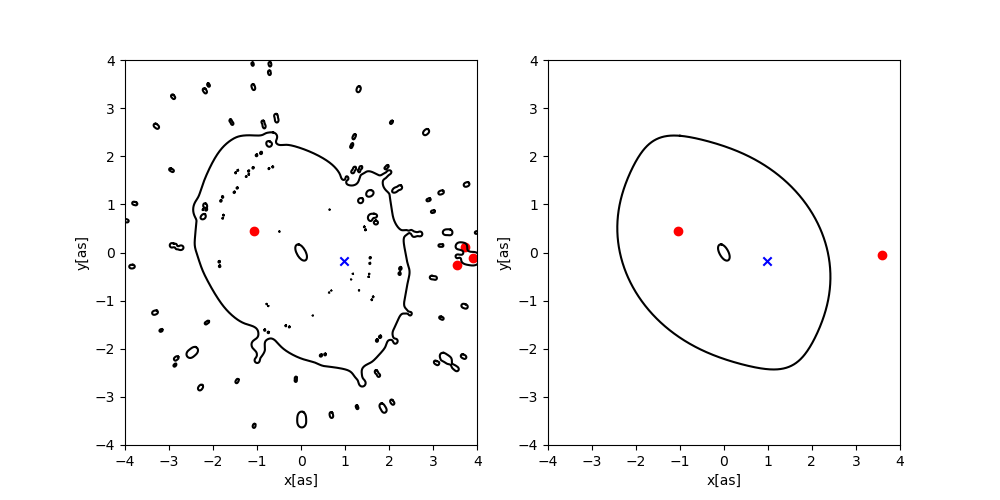}

\caption{Image number anomalies for 3 (top) and 4 (buttom) images induced from double lens identified easily from shape configurations. From left to right, 1st and 3rd panel show 3 (top) or 4 (buttom) images observed. 2nd and 4th panel show original double lens systems accordingly. Source positions are indicated by blue crosses, lensed images by red dots, and critical curves by black lines.}\label{direct}
\end{figure*}

A critical challenge in image-number anomaly analysis is the interpretation of image configurations. The analysis of the constraint results utilizes the known original image numbers of the investigated lensing systems to identify image-number anomalies. In the context of real survey data, a classification scheme is necessary to distinguish between lens systems with normal image numbers and with anomalous image numbers. We investigate two classes based on the number of images: systems with three and systems with four image numbers. Systems exhibiting more than four image numbers are categorically considered anomalous. The observation of three images (a triplet) presents an ambiguity: it could indicate the presence of a third image beyond the standard double lens, or it could represent a degenerate quadrupole system where one image of a theoretical quadruplet has been suppressed. This degeneracy extends to four-image systems (quadruplets), which can be unambiguously produced by a quadrupole lens but may also be generated under specific conditions by a double lens.

\begin{figure*}
\includegraphics[scale=0.4]{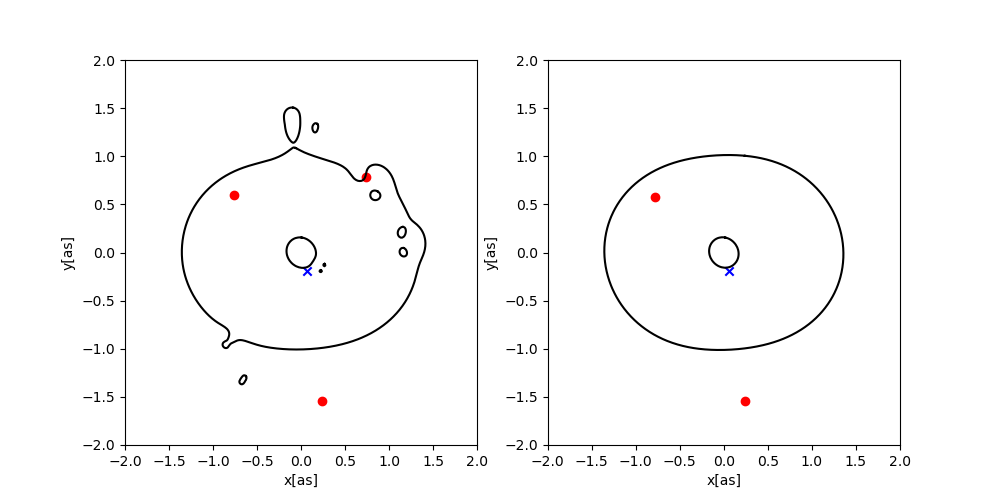}
\includegraphics[scale=0.4]{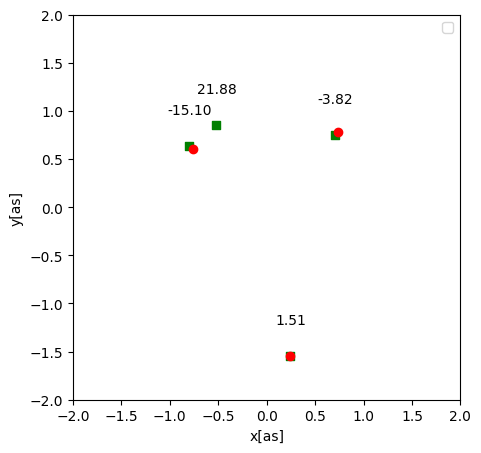}

\includegraphics[scale=0.4]{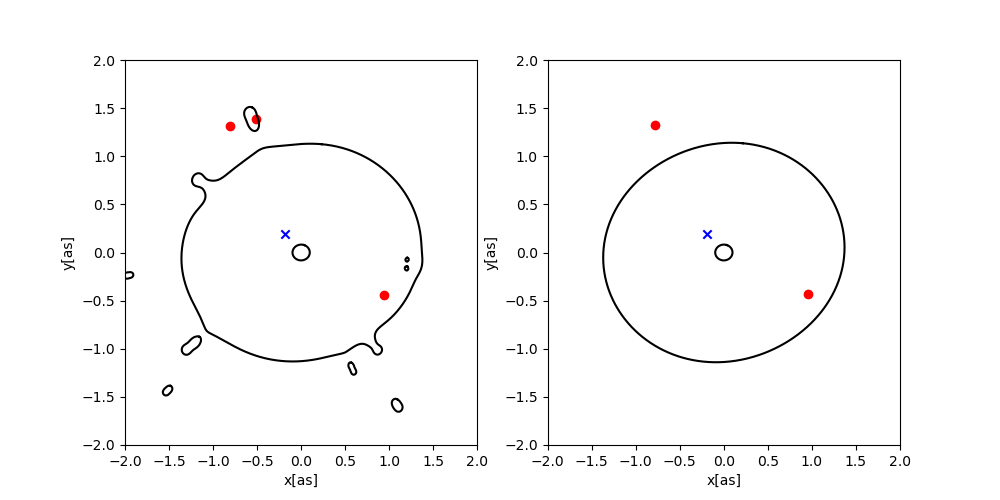}
\includegraphics[scale=0.4]{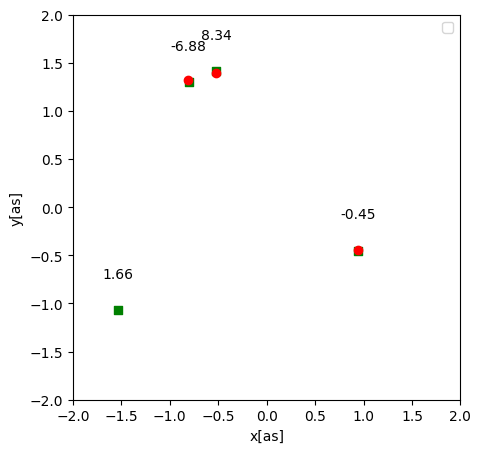}

\caption{
Image number anomalies (3 images induced from a double lens) identified through our fitting procedure. The left panels (top and bottom) display the observed image configurations. The middle panels show the corresponding original double-lens systems. The right panels present the fitted results where green boxes highlight the quadruple-image configurations generated from fitted parameters forming a cusp (top) and a fold (bottom), respectively. The values for each image stand for magnifications. Observed image positions are marked by red dots, while the source positions are indicated by blue crosses. Critical curves of the lens models are shown as black lines.}\label{fit1}
\end{figure*}

\begin{figure*}
\includegraphics[scale=0.4]{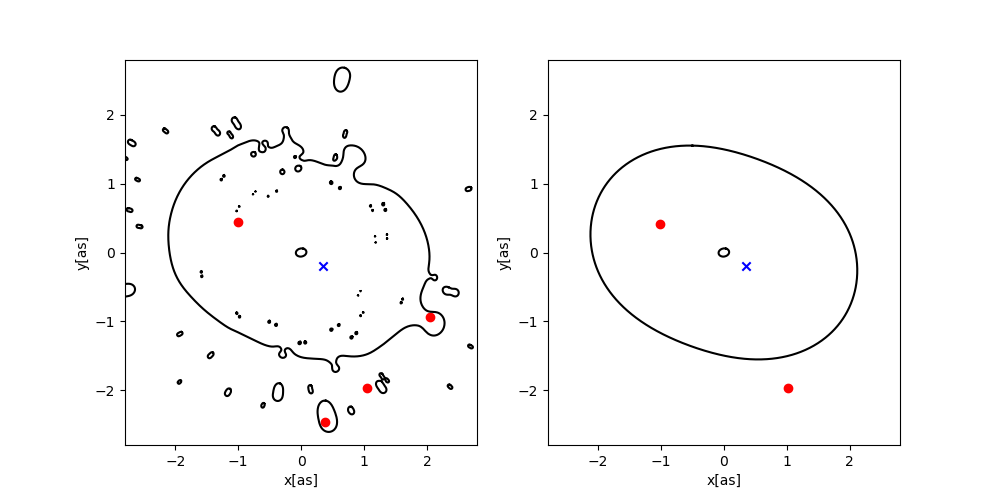}
\includegraphics[scale=0.4]{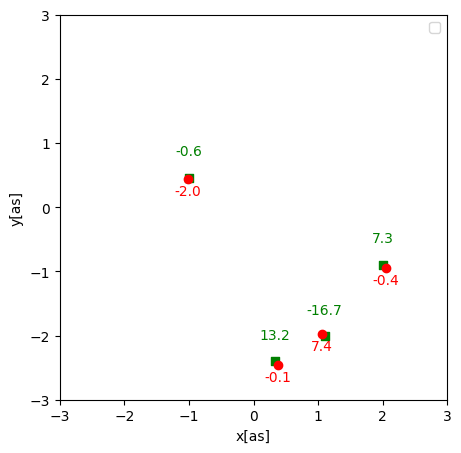}

\caption{An image number anomaly (4 images induced from a double lens) identified by fitting procedure presented. left panel shows 4 images observed. Middle panel shows the original double lens system. Right panel shows the fitted result of the cusp configuration (highlighted by green boxes). The values for each image stand for magnifications. Image positions (red dots), the source position (blue cross), critical curves (black lines), and a comparison of observed (red) and fitted (green) flux values are indicated.}\label{fit2}
\end{figure*}

We illustrate our identification methods through several representative cases. In Cases 1-4, the image configurations generated by PBH substructure, as depicted in Fig. \ref{direct}, exhibit anomalous morphologies that are atypical of standard lens systems. These configurations are readily identifiable as image number anomalies through direct visual inspection. Cases 5-7: As illustrated in Figs. \ref{fit1}, \ref{fit2}, discrimination between the cases proves challenging. To resolve this, a fitting method is further employed for differentiation. We note that image-number anomalies are rare in the FDM scenario involving double-lens systems and we restrict our analysis to the PBH scenario.

We now describe our fitting procedure. A composite lens model consisting of an elliptical NFW profile and an elliptical Hernquist profile is adopted. For simplicity, the lens center is fixed at the origin. External shear is neglected, and we assume that the structural parameters of the Hernquist profile are constrained by independent observations. Consequently, the free parameters in our model include four parameters of the elliptical NFW profile, one parameter of mass-to-light ratio for the baryonic component, and two parameters describing the source position. We employ the \texttt{emcee} \cite{2013ascl.soft03002F} package to sample the posterior probability distribution of these parameters using the observed positions of three (or four) lensed images. The goodness of fit is evaluated by the $\chi^2$ statistic, defined as
\begin{equation}\label{equation:error}
\chi^2 = \sum_{i=1}^{N} \frac{\left|\mathbf{x}_i^{\mathrm{obs}} - \mathbf{x}_i(\boldsymbol{\Theta})\right|^2}{\sigma_i^2},
\end{equation}
where $\mathbf{x}_i^{\mathrm{obs}}$ are the observed positions of the $N = 3$ or $4$ images, $\mathbf{x}_i(\boldsymbol{\Theta})$ are the corresponding image positions predicted by the composite model, and $\sigma_i$ denotes the positional uncertainty, which we set to $0.01''$ for all images. The posterior distribution is explored using Markov Chain Monte Carlo (MCMC) sampling with uniform priors on all model parameters $\boldsymbol{\Theta}$. We note that the results should be interpreted primarily as a consistency test of the model rather than a statistically rigorous parameter determination.

As illustrated in Fig.~\ref{fit1}, the lens model, constrained by the fitted parameters, predicts four lensed images, indicated by green boxes. The numerical values associated with each image denote the corresponding magnifications. Based on the flux comparison shown in the upper-right panel (e.g., a flux of 21.88 versus 1.51), the fourth image should, in principle, be detectable in the observational data. However, this image is not observed. We therefore attribute its absence to a image-number anomaly from a double-lens system.

When the source is positioned in the vicinity of a cusp of the caustic, three images, denoted A, B, and C, emerge in close proximity. According to standard gravitational lensing theory, their magnifications are expected to satisfy the cusp relation:
\begin{equation}
R_{\text{cusp}} = \frac{|\mu_A| - |\mu_B| + |\mu_C|}{|\mu_A| + |\mu_B| + |\mu_C|} \approx 0.
\end{equation}
However, the observed image fluxes (red numbers) in Fig.~\ref{fit2} deviate from this prediction, thereby violating the cusp relation. We take this lens system as an image number anomaly from a double-lens system similarly.

Microlensing induces only a finite flux variation, insufficient to explain the disappearance of an image with a large magnification, such as 21.88 in Fig.~\ref{fit1}. The potential contamination from stellar microlensing in the above flux ratio analyses can also be effectively mitigated by adopting observational strategies that target spatially extended emission regions. In addition, high-resolution facilities such as the Hubble Space Telescope (HST) and JWST can directly probe the narrow-line region (NLR) and the warm dust emission, both of which possess sufficient spatial extent to remain largely insensitive to stellar microlensing.

For lens systems that prove challenging to identify through the aforementioned methods, effective recognition remains achievable by integrating supplementary data, such as host galaxy light arcs and time delays. Even if some lens systems could not be definitively classified as anomalies, we are able to place conservative constraints on the properties of dark matter substructures. As machine learning gains widespread adoption, we can also employ machine learning algorithms  to identify such lens systems in the future.

\section{Summaries and discussion} \label{Sec V}

Anomalies in image position, flux ratio, and time delay have traditionally served as primary observables for investigating dark matter substructure. With the advancements in both spatial and temporal resolution of telescopes, we propose a new diagnostic, the image number anomaly, for the detection of dark matter substructure
We construct lens systems to demonstrate that dark matter substructures, specifically, PBHs and FDM, can induce image number anomalies. Furthermore, we show that such anomalies are more readily observable at higher angular resolutions. Utilizing a null result for image-number anomalies from 3,500 strong lensing systems, we derive constraints on PBHs and FDM. Our 95\% confidence-level constraints on the PBH fraction, $ f_{\mathrm{PBH}} $, are $ \lesssim 0.125\% $, $ 0.08\% $, and $ 0.04\% $ for PBH masses of $ 10^{7} $--$ 10^{9}~M_{\odot} $, achieved at angular resolutions of $ 0.1'' $, $ 0.05'' $, and $ 0.01'' $, respectively. In the FDM paradigm, we exclude particles with masses below $ 0.4 $, $ 0.6 $, and $ 3.5 \times 10^{-22} \ \mathrm{eV} $ for the same respective resolutions. We also constrain PBH abundances as low as $ 0.9\% $ at a resolution of $ 0.5'' $ for LSST observations. In addition to PBHs and FDM, we have tested CDM-subhalo scenario as well. In such case, the perturbation is relatively weak to generate enough image-number-anomaly systems. The reason could be that CDM subhalos have extended profiles compared to point-mass lenses. For the same mass, extended lenses are weaker in splitting images.

We further examine specific cases in our simulations. Beyond image number anomalies that are readily identifiable, the fit method can be employed to determine whether a lens system exhibits anomalous image numbers. Microlensing is not considered. First, it cannot explain the absence of detected large-flux images in certain cases. Second, observational strategies can mitigate stellar microlensing contamination in flux ratio analyses. Radio observations are ideal because quasar radio-emitting regions are larger than the microlens Einstein radius. High-resolution observations of extended features like the narrow-line region and warm dust with HST or JWST are also minimally affected. Moreover, we also could identify anomalies through the integration of host galaxy light arcs, time delays, and related metrics. In addition, machine learning techniques may be applied for detection in future work. Although some lens systems remain challenging to classify, they constitute a negligible fraction and do not impact our statistical conclusions.

Though we take PBHs and FDM as examples, our method is not limited to them. One can apply it for the perturbation by any dark matter candidates: CDM subhalos, PBHs, Warm dark matter, Self-interacting dark matter, Fuzzy dark matter etc. Note among these models, only PBHs can have a very large mass range, others usually correspond to millilensing range. 

As the catalog of strong gravitational lensing systems continues to grow, driven by advances in observational surveys and instrumentation, the development of diverse and robust methodological frameworks is increasingly imperative to fully leverage the scientific potential of forthcoming large-scale datasets. In this context, the image-number anomaly method introduced in this work offers a promising avenue for probing the small-scale structure of dark matter, thereby contributing to a deeper understanding of its fundamental properties.

\section*{Acknowledgments}
This work was supported by National Key R\&D Program of China (No. 2024YFC2207400) and National Natural Science Foundation of China (No. 12222302).

\nocite{*}

\bibliography{apssamp}% Produces the bibliography via BibTeX.

%apsrev4-2.bst 2019-01-14 (MD) hand-edited version of apsrev4-1.bst
%Control: key (0)
%Control: author (8) initials jnrlst
%Control: editor formatted (1) identically to author
%Control: production of article title (0) allowed
%Control: page (0) single
%Control: year (1) truncated
%Control: production of eprint (0) enabled
\providecommand{\noopsort}[1]{}\providecommand{\singleletter}[1]{#1}%
\begin{thebibliography}{43}%
\makeatletter
\providecommand \@ifxundefined [1]{%
 \@ifx{#1\undefined}
}%
\providecommand \@ifnum [1]{%
 \ifnum #1\expandafter \@firstoftwo
 \else \expandafter \@secondoftwo
 \fi
}%
\providecommand \@ifx [1]{%
 \ifx #1\expandafter \@firstoftwo
 \else \expandafter \@secondoftwo
 \fi
}%
\providecommand \natexlab [1]{#1}%
\providecommand \enquote  [1]{``#1''}%
\providecommand \bibnamefont  [1]{#1}%
\providecommand \bibfnamefont [1]{#1}%
\providecommand \citenamefont [1]{#1}%
\providecommand \href@noop [0]{\@secondoftwo}%
\providecommand \href [0]{\begingroup \@sanitize@url \@href}%
\providecommand \@href[1]{\@@startlink{#1}\@@href}%
\providecommand \@@href[1]{\endgroup#1\@@endlink}%
\providecommand \@sanitize@url [0]{\catcode `\\12\catcode `\$12\catcode `\&12\catcode `\#12\catcode `\^12\catcode `\_12\catcode `\%12\relax}%
\providecommand \@@startlink[1]{}%
\providecommand \@@endlink[0]{}%
\providecommand \url  [0]{\begingroup\@sanitize@url \@url }%
\providecommand \@url [1]{\endgroup\@href {#1}{\urlprefix }}%
\providecommand \urlprefix  [0]{URL }%
\providecommand \Eprint [0]{\href }%
\providecommand \doibase [0]{https://doi.org/}%
\providecommand \selectlanguage [0]{\@gobble}%
\providecommand \bibinfo  [0]{\@secondoftwo}%
\providecommand \bibfield  [0]{\@secondoftwo}%
\providecommand \translation [1]{[#1]}%
\providecommand \BibitemOpen [0]{}%
\providecommand \bibitemStop [0]{}%
\providecommand \bibitemNoStop [0]{.\EOS\space}%
\providecommand \EOS [0]{\spacefactor3000\relax}%
\providecommand \BibitemShut  [1]{\csname bibitem#1\endcsname}%
\let\auto@bib@innerbib\@empty
%</preamble>
\bibitem [{\citenamefont {{Rubin}}\ and\ \citenamefont {{Ford}}(1970)}]{Rubin1978}%
  \BibitemOpen
  \bibfield  {author} {\bibinfo {author} {\bibfnamefont {V.~C.}\ \bibnamefont {{Rubin}}}\ and\ \bibinfo {author} {\bibfnamefont {W.~K.}\ \bibnamefont {{Ford}}, \bibfnamefont {Jr.}},\ }\bibfield  {title} {\bibinfo {title} {{Rotation of the Andromeda Nebula from a Spectroscopic Survey of Emission Regions}},\ }\href {https://doi.org/10.1086/150317} {\bibfield  {journal} {\bibinfo  {journal} {The Astrophysical Journal}\ }\textbf {\bibinfo {volume} {159}},\ \bibinfo {pages} {379} (\bibinfo {year} {1970})}\BibitemShut {NoStop}%
\bibitem [{\citenamefont {{Clowe}}\ \emph {et~al.}(2006)\citenamefont {{Clowe}}, \citenamefont {{Brada{\v{c}}}}, \citenamefont {{Gonzalez}}, \citenamefont {{Markevitch}}, \citenamefont {{Randall}}, \citenamefont {{Jones}},\ and\ \citenamefont {{Zaritsky}}}]{Clowe2006}%
  \BibitemOpen
  \bibfield  {author} {\bibinfo {author} {\bibfnamefont {D.}~\bibnamefont {{Clowe}}}, \bibinfo {author} {\bibfnamefont {M.}~\bibnamefont {{Brada{\v{c}}}}}, \bibinfo {author} {\bibfnamefont {A.~H.}\ \bibnamefont {{Gonzalez}}}, \bibinfo {author} {\bibfnamefont {M.}~\bibnamefont {{Markevitch}}}, \bibinfo {author} {\bibfnamefont {S.~W.}\ \bibnamefont {{Randall}}}, \bibinfo {author} {\bibfnamefont {C.}~\bibnamefont {{Jones}}},\ and\ \bibinfo {author} {\bibfnamefont {D.}~\bibnamefont {{Zaritsky}}},\ }\bibfield  {title} {\bibinfo {title} {{A Direct Empirical Proof of the Existence of Dark Matter}},\ }\href {https://doi.org/10.1086/508162} {\bibfield  {journal} {\bibinfo  {journal} {The Astrophysical Journal Letters}\ }\textbf {\bibinfo {volume} {648}},\ \bibinfo {pages} {L109} (\bibinfo {year} {2006})},\ \Eprint {https://arxiv.org/abs/astro-ph/0608407} {arXiv:astro-ph/0608407 [astro-ph]} \BibitemShut {NoStop}%
\bibitem [{\citenamefont {{Blumenthal}}\ \emph {et~al.}(1984)\citenamefont {{Blumenthal}}, \citenamefont {{Faber}}, \citenamefont {{Primack}},\ and\ \citenamefont {{Rees}}}]{Blumenthal1984}%
  \BibitemOpen
  \bibfield  {author} {\bibinfo {author} {\bibfnamefont {G.~R.}\ \bibnamefont {{Blumenthal}}}, \bibinfo {author} {\bibfnamefont {S.~M.}\ \bibnamefont {{Faber}}}, \bibinfo {author} {\bibfnamefont {J.~R.}\ \bibnamefont {{Primack}}},\ and\ \bibinfo {author} {\bibfnamefont {M.~J.}\ \bibnamefont {{Rees}}},\ }\bibfield  {title} {\bibinfo {title} {{Formation of galaxies and large-scale structure with cold dark matter.}},\ }\href {https://doi.org/10.1038/311517a0} {\bibfield  {journal} {\bibinfo  {journal} {Nature}\ }\textbf {\bibinfo {volume} {311}},\ \bibinfo {pages} {517} (\bibinfo {year} {1984})}\BibitemShut {NoStop}%
\bibitem [{\citenamefont {{Klypin}}\ \emph {et~al.}(1999)\citenamefont {{Klypin}}, \citenamefont {{Kravtsov}}, \citenamefont {{Valenzuela}},\ and\ \citenamefont {{Prada}}}]{Klypin1999}%
  \BibitemOpen
  \bibfield  {author} {\bibinfo {author} {\bibfnamefont {A.}~\bibnamefont {{Klypin}}}, \bibinfo {author} {\bibfnamefont {A.~V.}\ \bibnamefont {{Kravtsov}}}, \bibinfo {author} {\bibfnamefont {O.}~\bibnamefont {{Valenzuela}}},\ and\ \bibinfo {author} {\bibfnamefont {F.}~\bibnamefont {{Prada}}},\ }\bibfield  {title} {\bibinfo {title} {{Where Are the Missing Galactic Satellites?}},\ }\href {https://doi.org/10.1086/307643} {\bibfield  {journal} {\bibinfo  {journal} {The Astrophysical Journal}\ }\textbf {\bibinfo {volume} {522}},\ \bibinfo {pages} {82} (\bibinfo {year} {1999})},\ \Eprint {https://arxiv.org/abs/astro-ph/9901240} {arXiv:astro-ph/9901240 [astro-ph]} \BibitemShut {NoStop}%
\bibitem [{\citenamefont {{Moore}}\ \emph {et~al.}(1999)\citenamefont {{Moore}}, \citenamefont {{Ghigna}}, \citenamefont {{Governato}}, \citenamefont {{Lake}}, \citenamefont {{Quinn}}, \citenamefont {{Stadel}},\ and\ \citenamefont {{Tozzi}}}]{Moore1999}%
  \BibitemOpen
  \bibfield  {author} {\bibinfo {author} {\bibfnamefont {B.}~\bibnamefont {{Moore}}}, \bibinfo {author} {\bibfnamefont {S.}~\bibnamefont {{Ghigna}}}, \bibinfo {author} {\bibfnamefont {F.}~\bibnamefont {{Governato}}}, \bibinfo {author} {\bibfnamefont {G.}~\bibnamefont {{Lake}}}, \bibinfo {author} {\bibfnamefont {T.}~\bibnamefont {{Quinn}}}, \bibinfo {author} {\bibfnamefont {J.}~\bibnamefont {{Stadel}}},\ and\ \bibinfo {author} {\bibfnamefont {P.}~\bibnamefont {{Tozzi}}},\ }\bibfield  {title} {\bibinfo {title} {{Dark Matter Substructure within Galactic Halos}},\ }\href {https://doi.org/10.1086/312287} {\bibfield  {journal} {\bibinfo  {journal} {The Astrophysical Journal Letters}\ }\textbf {\bibinfo {volume} {524}},\ \bibinfo {pages} {L19} (\bibinfo {year} {1999})},\ \Eprint {https://arxiv.org/abs/astro-ph/9907411} {arXiv:astro-ph/9907411 [astro-ph]} \BibitemShut {NoStop}%
\bibitem [{\citenamefont {{de Blok}}(2010)}]{deBlok2010}%
  \BibitemOpen
  \bibfield  {author} {\bibinfo {author} {\bibfnamefont {W.~J.~G.}\ \bibnamefont {{de Blok}}},\ }\bibfield  {title} {\bibinfo {title} {{The Core-Cusp Problem}},\ }\href {https://doi.org/10.1155/2010/789293} {\bibfield  {journal} {\bibinfo  {journal} {Advances in Astronomy}\ }\textbf {\bibinfo {volume} {2010}},\ \bibinfo {eid} {789293} (\bibinfo {year} {2010})},\ \Eprint {https://arxiv.org/abs/0910.3538} {arXiv:0910.3538 [astro-ph.CO]} \BibitemShut {NoStop}%
\bibitem [{\citenamefont {{Carr}}\ and\ \citenamefont {{K{\"u}hnel}}(2020)}]{Carr2020}%
  \BibitemOpen
  \bibfield  {author} {\bibinfo {author} {\bibfnamefont {B.}~\bibnamefont {{Carr}}}\ and\ \bibinfo {author} {\bibfnamefont {F.}~\bibnamefont {{K{\"u}hnel}}},\ }\bibfield  {title} {\bibinfo {title} {{Primordial Black Holes as Dark Matter: Recent Developments}},\ }\href {https://doi.org/10.1146/annurev-nucl-050520-125911} {\bibfield  {journal} {\bibinfo  {journal} {Annual Review of Nuclear and Particle Science}\ }\textbf {\bibinfo {volume} {70}},\ \bibinfo {pages} {355} (\bibinfo {year} {2020})},\ \Eprint {https://arxiv.org/abs/2006.02838} {arXiv:2006.02838 [astro-ph.CO]} \BibitemShut {NoStop}%
\bibitem [{\citenamefont {{Hu}}\ \emph {et~al.}(2000)\citenamefont {{Hu}}, \citenamefont {{Barkana}},\ and\ \citenamefont {{Gruzinov}}}]{Hu2000}%
  \BibitemOpen
  \bibfield  {author} {\bibinfo {author} {\bibfnamefont {W.}~\bibnamefont {{Hu}}}, \bibinfo {author} {\bibfnamefont {R.}~\bibnamefont {{Barkana}}},\ and\ \bibinfo {author} {\bibfnamefont {A.}~\bibnamefont {{Gruzinov}}},\ }\bibfield  {title} {\bibinfo {title} {{Fuzzy Cold Dark Matter: The Wave Properties of Ultralight Particles}},\ }\href {https://doi.org/10.1103/PhysRevLett.85.1158} {\bibfield  {journal} {\bibinfo  {journal} {Physical Review Letters}\ }\textbf {\bibinfo {volume} {85}},\ \bibinfo {pages} {1158} (\bibinfo {year} {2000})},\ \Eprint {https://arxiv.org/abs/astro-ph/0003365} {arXiv:astro-ph/0003365 [astro-ph]} \BibitemShut {NoStop}%
\bibitem [{\citenamefont {{Hui}}\ \emph {et~al.}(2017)\citenamefont {{Hui}}, \citenamefont {{Ostriker}}, \citenamefont {{Tremaine}},\ and\ \citenamefont {{Witten}}}]{Hui2017}%
  \BibitemOpen
  \bibfield  {author} {\bibinfo {author} {\bibfnamefont {L.}~\bibnamefont {{Hui}}}, \bibinfo {author} {\bibfnamefont {J.~P.}\ \bibnamefont {{Ostriker}}}, \bibinfo {author} {\bibfnamefont {S.}~\bibnamefont {{Tremaine}}},\ and\ \bibinfo {author} {\bibfnamefont {E.}~\bibnamefont {{Witten}}},\ }\bibfield  {title} {\bibinfo {title} {{Ultralight scalars as cosmological dark matter}},\ }\href {https://doi.org/10.1103/PhysRevD.95.043541} {\bibfield  {journal} {\bibinfo  {journal} {Physical Review D}\ }\textbf {\bibinfo {volume} {95}},\ \bibinfo {eid} {043541} (\bibinfo {year} {2017})},\ \Eprint {https://arxiv.org/abs/1610.08297} {arXiv:1610.08297 [astro-ph.CO]} \BibitemShut {NoStop}%
\bibitem [{\citenamefont {{Kelly}}\ \emph {et~al.}(2018)\citenamefont {{Kelly}}, \citenamefont {{Diego}}, \citenamefont {{Rodney}}, \citenamefont {{Kaiser}}, \citenamefont {{Broadhurst}}, \citenamefont {{Zitrin}}, \citenamefont {{Treu}}, \citenamefont {{P{\'e}rez-Gonz{\'a}lez}}, \citenamefont {{Morishita}}, \citenamefont {{Jauzac}}, \citenamefont {{Selsing}}, \citenamefont {{Oguri}}, \citenamefont {{Pueyo}}, \citenamefont {{Ross}}, \citenamefont {{Filippenko}}, \citenamefont {{Smith}}, \citenamefont {{Hjorth}}, \citenamefont {{Cenko}}, \citenamefont {{Wang}}, \citenamefont {{Howell}}, \citenamefont {{Richard}}, \citenamefont {{Frye}}, \citenamefont {{Jha}}, \citenamefont {{Foley}}, \citenamefont {{Norman}}, \citenamefont {{Bradac}}, \citenamefont {{Zheng}}, \citenamefont {{Brammer}}, \citenamefont {{Benito}}, \citenamefont {{Cava}}, \citenamefont {{Christensen}}, \citenamefont {{de Mink}}, \citenamefont {{Graur}}, \citenamefont {{Grillo}}, \citenamefont {{Kawamata}}, \citenamefont {{Kneib}}, \citenamefont
  {{Matheson}}, \citenamefont {{McCully}}, \citenamefont {{Nonino}}, \citenamefont {{P{\'e}rez-Fournon}}, \citenamefont {{Riess}}, \citenamefont {{Rosati}}, \citenamefont {{Schmidt}}, \citenamefont {{Sharon}},\ and\ \citenamefont {{Weiner}}}]{2018NatAs...2..334K}%
  \BibitemOpen
  \bibfield  {author} {\bibinfo {author} {\bibfnamefont {P.~L.}\ \bibnamefont {{Kelly}}}, \bibinfo {author} {\bibfnamefont {J.~M.}\ \bibnamefont {{Diego}}}, \bibinfo {author} {\bibfnamefont {S.}~\bibnamefont {{Rodney}}}, \bibinfo {author} {\bibfnamefont {N.}~\bibnamefont {{Kaiser}}}, \bibinfo {author} {\bibfnamefont {T.}~\bibnamefont {{Broadhurst}}}, \bibinfo {author} {\bibfnamefont {A.}~\bibnamefont {{Zitrin}}}, \bibinfo {author} {\bibfnamefont {T.}~\bibnamefont {{Treu}}}, \bibinfo {author} {\bibfnamefont {P.~G.}\ \bibnamefont {{P{\'e}rez-Gonz{\'a}lez}}}, \bibinfo {author} {\bibfnamefont {T.}~\bibnamefont {{Morishita}}}, \bibinfo {author} {\bibfnamefont {M.}~\bibnamefont {{Jauzac}}}, \bibinfo {author} {\bibfnamefont {J.}~\bibnamefont {{Selsing}}}, \bibinfo {author} {\bibfnamefont {M.}~\bibnamefont {{Oguri}}}, \bibinfo {author} {\bibfnamefont {L.}~\bibnamefont {{Pueyo}}}, \bibinfo {author} {\bibfnamefont {T.~W.}\ \bibnamefont {{Ross}}}, \bibinfo {author} {\bibfnamefont {A.~V.}\ \bibnamefont {{Filippenko}}},
  \bibinfo {author} {\bibfnamefont {N.}~\bibnamefont {{Smith}}}, \bibinfo {author} {\bibfnamefont {J.}~\bibnamefont {{Hjorth}}}, \bibinfo {author} {\bibfnamefont {S.~B.}\ \bibnamefont {{Cenko}}}, \bibinfo {author} {\bibfnamefont {X.}~\bibnamefont {{Wang}}}, \bibinfo {author} {\bibfnamefont {D.~A.}\ \bibnamefont {{Howell}}}, \bibinfo {author} {\bibfnamefont {J.}~\bibnamefont {{Richard}}}, \bibinfo {author} {\bibfnamefont {B.~L.}\ \bibnamefont {{Frye}}}, \bibinfo {author} {\bibfnamefont {S.~W.}\ \bibnamefont {{Jha}}}, \bibinfo {author} {\bibfnamefont {R.~J.}\ \bibnamefont {{Foley}}}, \bibinfo {author} {\bibfnamefont {C.}~\bibnamefont {{Norman}}}, \bibinfo {author} {\bibfnamefont {M.}~\bibnamefont {{Bradac}}}, \bibinfo {author} {\bibfnamefont {W.}~\bibnamefont {{Zheng}}}, \bibinfo {author} {\bibfnamefont {G.}~\bibnamefont {{Brammer}}}, \bibinfo {author} {\bibfnamefont {A.~M.}\ \bibnamefont {{Benito}}}, \bibinfo {author} {\bibfnamefont {A.}~\bibnamefont {{Cava}}}, \bibinfo {author} {\bibfnamefont
  {L.}~\bibnamefont {{Christensen}}}, \bibinfo {author} {\bibfnamefont {S.~E.}\ \bibnamefont {{de Mink}}}, \bibinfo {author} {\bibfnamefont {O.}~\bibnamefont {{Graur}}}, \bibinfo {author} {\bibfnamefont {C.}~\bibnamefont {{Grillo}}}, \bibinfo {author} {\bibfnamefont {R.}~\bibnamefont {{Kawamata}}}, \bibinfo {author} {\bibfnamefont {J.-P.}\ \bibnamefont {{Kneib}}}, \bibinfo {author} {\bibfnamefont {T.}~\bibnamefont {{Matheson}}}, \bibinfo {author} {\bibfnamefont {C.}~\bibnamefont {{McCully}}}, \bibinfo {author} {\bibfnamefont {M.}~\bibnamefont {{Nonino}}}, \bibinfo {author} {\bibfnamefont {I.}~\bibnamefont {{P{\'e}rez-Fournon}}}, \bibinfo {author} {\bibfnamefont {A.~G.}\ \bibnamefont {{Riess}}}, \bibinfo {author} {\bibfnamefont {P.}~\bibnamefont {{Rosati}}}, \bibinfo {author} {\bibfnamefont {K.~B.}\ \bibnamefont {{Schmidt}}}, \bibinfo {author} {\bibfnamefont {K.}~\bibnamefont {{Sharon}}},\ and\ \bibinfo {author} {\bibfnamefont {B.~J.}\ \bibnamefont {{Weiner}}},\ }\bibfield  {title} {\bibinfo {title} {{Extreme
  magnification of an individual star at redshift 1.5 by a galaxy-cluster lens}},\ }\href {https://doi.org/10.1038/s41550-018-0430-3} {\bibfield  {journal} {\bibinfo  {journal} {Nature Astronomy}\ }\textbf {\bibinfo {volume} {2}},\ \bibinfo {pages} {334} (\bibinfo {year} {2018})},\ \Eprint {https://arxiv.org/abs/1706.10279} {arXiv:1706.10279 [astro-ph.GA]} \BibitemShut {NoStop}%
\bibitem [{\citenamefont {{Oguri}}\ \emph {et~al.}(2018)\citenamefont {{Oguri}}, \citenamefont {{Diego}}, \citenamefont {{Kaiser}}, \citenamefont {{Kelly}},\ and\ \citenamefont {{Broadhurst}}}]{2018PhRvD..97b3518O}%
  \BibitemOpen
  \bibfield  {author} {\bibinfo {author} {\bibfnamefont {M.}~\bibnamefont {{Oguri}}}, \bibinfo {author} {\bibfnamefont {J.~M.}\ \bibnamefont {{Diego}}}, \bibinfo {author} {\bibfnamefont {N.}~\bibnamefont {{Kaiser}}}, \bibinfo {author} {\bibfnamefont {P.~L.}\ \bibnamefont {{Kelly}}},\ and\ \bibinfo {author} {\bibfnamefont {T.}~\bibnamefont {{Broadhurst}}},\ }\bibfield  {title} {\bibinfo {title} {{Understanding caustic crossings in giant arcs: Characteristic scales, event rates, and constraints on compact dark matter}},\ }\href {https://doi.org/10.1103/PhysRevD.97.023518} {\bibfield  {journal} {\bibinfo  {journal} {\prd}\ }\textbf {\bibinfo {volume} {97}},\ \bibinfo {eid} {023518} (\bibinfo {year} {2018})},\ \Eprint {https://arxiv.org/abs/1710.00148} {arXiv:1710.00148 [astro-ph.CO]} \BibitemShut {NoStop}%
\bibitem [{\citenamefont {{Diego}}\ \emph {et~al.}(2023)\citenamefont {{Diego}}, \citenamefont {{Sun}}, \citenamefont {{Yan}}, \citenamefont {{Furtak}}, \citenamefont {{Zackrisson}}, \citenamefont {{Dai}}, \citenamefont {{Kelly}}, \citenamefont {{Nonino}}, \citenamefont {{Adams}}, \citenamefont {{Meena}}, \citenamefont {{Willner}}, \citenamefont {{Zitrin}}, \citenamefont {{Cohen}}, \citenamefont {{D'Silva}}, \citenamefont {{Jansen}}, \citenamefont {{Summers}}, \citenamefont {{Windhorst}}, \citenamefont {{Coe}}, \citenamefont {{Conselice}}, \citenamefont {{Driver}}, \citenamefont {{Frye}}, \citenamefont {{Grogin}}, \citenamefont {{Koekemoer}}, \citenamefont {{Marshall}}, \citenamefont {{Pirzkal}}, \citenamefont {{Robotham}}, \citenamefont {{Rutkowski}}, \citenamefont {{Ryan}}, \citenamefont {{Tompkins}}, \citenamefont {{Willmer}},\ and\ \citenamefont {{Bhatawdekar}}}]{2023A&A...679A..31D}%
  \BibitemOpen
  \bibfield  {author} {\bibinfo {author} {\bibfnamefont {J.~M.}\ \bibnamefont {{Diego}}}, \bibinfo {author} {\bibfnamefont {B.}~\bibnamefont {{Sun}}}, \bibinfo {author} {\bibfnamefont {H.}~\bibnamefont {{Yan}}}, \bibinfo {author} {\bibfnamefont {L.~J.}\ \bibnamefont {{Furtak}}}, \bibinfo {author} {\bibfnamefont {E.}~\bibnamefont {{Zackrisson}}}, \bibinfo {author} {\bibfnamefont {L.}~\bibnamefont {{Dai}}}, \bibinfo {author} {\bibfnamefont {P.}~\bibnamefont {{Kelly}}}, \bibinfo {author} {\bibfnamefont {M.}~\bibnamefont {{Nonino}}}, \bibinfo {author} {\bibfnamefont {N.}~\bibnamefont {{Adams}}}, \bibinfo {author} {\bibfnamefont {A.~K.}\ \bibnamefont {{Meena}}}, \bibinfo {author} {\bibfnamefont {S.~P.}\ \bibnamefont {{Willner}}}, \bibinfo {author} {\bibfnamefont {A.}~\bibnamefont {{Zitrin}}}, \bibinfo {author} {\bibfnamefont {S.~H.}\ \bibnamefont {{Cohen}}}, \bibinfo {author} {\bibfnamefont {J.~C.~J.}\ \bibnamefont {{D'Silva}}}, \bibinfo {author} {\bibfnamefont {R.~A.}\ \bibnamefont {{Jansen}}}, \bibinfo {author}
  {\bibfnamefont {J.}~\bibnamefont {{Summers}}}, \bibinfo {author} {\bibfnamefont {R.~A.}\ \bibnamefont {{Windhorst}}}, \bibinfo {author} {\bibfnamefont {D.}~\bibnamefont {{Coe}}}, \bibinfo {author} {\bibfnamefont {C.~J.}\ \bibnamefont {{Conselice}}}, \bibinfo {author} {\bibfnamefont {S.~P.}\ \bibnamefont {{Driver}}}, \bibinfo {author} {\bibfnamefont {B.}~\bibnamefont {{Frye}}}, \bibinfo {author} {\bibfnamefont {N.~A.}\ \bibnamefont {{Grogin}}}, \bibinfo {author} {\bibfnamefont {A.~M.}\ \bibnamefont {{Koekemoer}}}, \bibinfo {author} {\bibfnamefont {M.~A.}\ \bibnamefont {{Marshall}}}, \bibinfo {author} {\bibfnamefont {N.}~\bibnamefont {{Pirzkal}}}, \bibinfo {author} {\bibfnamefont {A.}~\bibnamefont {{Robotham}}}, \bibinfo {author} {\bibfnamefont {M.~J.}\ \bibnamefont {{Rutkowski}}}, \bibinfo {author} {\bibfnamefont {R.~E.}\ \bibnamefont {{Ryan}}}, \bibinfo {author} {\bibfnamefont {S.}~\bibnamefont {{Tompkins}}}, \bibinfo {author} {\bibfnamefont {C.~N.~A.}\ \bibnamefont {{Willmer}}},\ and\ \bibinfo {author}
  {\bibfnamefont {R.}~\bibnamefont {{Bhatawdekar}}},\ }\bibfield  {title} {\bibinfo {title} {{JWST's PEARLS: Mothra, a new kaiju star at z = 2.091 extremely magnified by MACS0416, and implications for dark matter models}},\ }\href {https://doi.org/10.1051/0004-6361/202347556} {\bibfield  {journal} {\bibinfo  {journal} {Astronomy \& Astrophysics}\ }\textbf {\bibinfo {volume} {679}},\ \bibinfo {eid} {A31} (\bibinfo {year} {2023})},\ \Eprint {https://arxiv.org/abs/2307.10363} {arXiv:2307.10363 [astro-ph.CO]} \BibitemShut {NoStop}%
\bibitem [{\citenamefont {{Amruth}}\ \emph {et~al.}(2023)\citenamefont {{Amruth}}, \citenamefont {{Broadhurst}}, \citenamefont {{Lim}}, \citenamefont {{Oguri}}, \citenamefont {{Smoot}}, \citenamefont {{Diego}}, \citenamefont {{Leung}}, \citenamefont {{Emami}}, \citenamefont {{Li}}, \citenamefont {{Chiueh}}, \citenamefont {{Schive}}, \citenamefont {{Yeung}},\ and\ \citenamefont {{Li}}}]{2023NatAs...7..736A}%
  \BibitemOpen
  \bibfield  {author} {\bibinfo {author} {\bibfnamefont {A.}~\bibnamefont {{Amruth}}}, \bibinfo {author} {\bibfnamefont {T.}~\bibnamefont {{Broadhurst}}}, \bibinfo {author} {\bibfnamefont {J.}~\bibnamefont {{Lim}}}, \bibinfo {author} {\bibfnamefont {M.}~\bibnamefont {{Oguri}}}, \bibinfo {author} {\bibfnamefont {G.~F.}\ \bibnamefont {{Smoot}}}, \bibinfo {author} {\bibfnamefont {J.~M.}\ \bibnamefont {{Diego}}}, \bibinfo {author} {\bibfnamefont {E.}~\bibnamefont {{Leung}}}, \bibinfo {author} {\bibfnamefont {R.}~\bibnamefont {{Emami}}}, \bibinfo {author} {\bibfnamefont {J.}~\bibnamefont {{Li}}}, \bibinfo {author} {\bibfnamefont {T.}~\bibnamefont {{Chiueh}}}, \bibinfo {author} {\bibfnamefont {H.-Y.}\ \bibnamefont {{Schive}}}, \bibinfo {author} {\bibfnamefont {M.~C.~H.}\ \bibnamefont {{Yeung}}},\ and\ \bibinfo {author} {\bibfnamefont {S.~K.}\ \bibnamefont {{Li}}},\ }\bibfield  {title} {\bibinfo {title} {{Einstein rings modulated by wavelike dark matter from anomalies in gravitationally lensed images}},\ }\href
  {https://doi.org/10.1038/s41550-023-01943-9} {\bibfield  {journal} {\bibinfo  {journal} {Nature Astronomy}\ }\textbf {\bibinfo {volume} {7}},\ \bibinfo {pages} {736} (\bibinfo {year} {2023})},\ \Eprint {https://arxiv.org/abs/2304.09895} {arXiv:2304.09895 [astro-ph.CO]} \BibitemShut {NoStop}%
\bibitem [{\citenamefont {{Bullock}}\ and\ \citenamefont {{Boylan-Kolchin}}(2017)}]{Bullock2017}%
  \BibitemOpen
  \bibfield  {author} {\bibinfo {author} {\bibfnamefont {J.~S.}\ \bibnamefont {{Bullock}}}\ and\ \bibinfo {author} {\bibfnamefont {M.}~\bibnamefont {{Boylan-Kolchin}}},\ }\bibfield  {title} {\bibinfo {title} {{Small-Scale Challenges to the {\ensuremath{\Lambda}}CDM Paradigm}},\ }\href {https://doi.org/10.1146/annurev-astro-091916-055313} {\bibfield  {journal} {\bibinfo  {journal} {Annual Review of Astronomy and Astrophysics}\ }\textbf {\bibinfo {volume} {55}},\ \bibinfo {pages} {343} (\bibinfo {year} {2017})},\ \Eprint {https://arxiv.org/abs/1707.04256} {arXiv:1707.04256 [astro-ph.CO]} \BibitemShut {NoStop}%
\bibitem [{\citenamefont {{Vegetti}}\ and\ \citenamefont {{Koopmans}}(2009)}]{Vegetti2009}%
  \BibitemOpen
  \bibfield  {author} {\bibinfo {author} {\bibfnamefont {S.}~\bibnamefont {{Vegetti}}}\ and\ \bibinfo {author} {\bibfnamefont {L.~V.~E.}\ \bibnamefont {{Koopmans}}},\ }\bibfield  {title} {\bibinfo {title} {{Bayesian strong gravitational-lens modelling on adaptive grids: objective detection of mass substructure in Galaxies}},\ }\href {https://doi.org/10.1111/j.1365-2966.2008.14005.x} {\bibfield  {journal} {\bibinfo  {journal} {Monthly Notices of the Royal Astronomical Society}\ }\textbf {\bibinfo {volume} {392}},\ \bibinfo {pages} {945} (\bibinfo {year} {2009})},\ \Eprint {https://arxiv.org/abs/0805.0201} {arXiv:0805.0201 [astro-ph]} \BibitemShut {NoStop}%
\bibitem [{\citenamefont {{Dai}}\ \emph {et~al.}(2018)\citenamefont {{Dai}}, \citenamefont {{Venumadhav}}, \citenamefont {{Kaurov}},\ and\ \citenamefont {{Miralda-Escud}}}]{2018ApJ...867...24D}%
  \BibitemOpen
  \bibfield  {author} {\bibinfo {author} {\bibfnamefont {L.}~\bibnamefont {{Dai}}}, \bibinfo {author} {\bibfnamefont {T.}~\bibnamefont {{Venumadhav}}}, \bibinfo {author} {\bibfnamefont {A.~A.}\ \bibnamefont {{Kaurov}}},\ and\ \bibinfo {author} {\bibfnamefont {J.}~\bibnamefont {{Miralda-Escud}}},\ }\bibfield  {title} {\bibinfo {title} {{Probing Dark Matter Subhalos in Galaxy Clusters Using Highly Magnified Stars}},\ }\href {https://doi.org/10.3847/1538-4357/aae478} {\bibfield  {journal} {\bibinfo  {journal} {\apj}\ }\textbf {\bibinfo {volume} {867}},\ \bibinfo {eid} {24} (\bibinfo {year} {2018})},\ \Eprint {https://arxiv.org/abs/1804.03149} {arXiv:1804.03149 [astro-ph.CO]} \BibitemShut {NoStop}%
\bibitem [{\citenamefont {{Williams}}\ \emph {et~al.}(2024)\citenamefont {{Williams}}, \citenamefont {{Kelly}}, \citenamefont {{Treu}}, \citenamefont {{Amruth}}, \citenamefont {{Diego}}, \citenamefont {{Li}}, \citenamefont {{Meena}}, \citenamefont {{Zitrin}}, \citenamefont {{Broadhurst}},\ and\ \citenamefont {{Filippenko}}}]{2024ApJ...961..200W}%
  \BibitemOpen
  \bibfield  {author} {\bibinfo {author} {\bibfnamefont {L.~L.~R.}\ \bibnamefont {{Williams}}}, \bibinfo {author} {\bibfnamefont {P.~L.}\ \bibnamefont {{Kelly}}}, \bibinfo {author} {\bibfnamefont {T.}~\bibnamefont {{Treu}}}, \bibinfo {author} {\bibfnamefont {A.}~\bibnamefont {{Amruth}}}, \bibinfo {author} {\bibfnamefont {J.~M.}\ \bibnamefont {{Diego}}}, \bibinfo {author} {\bibfnamefont {S.~K.}\ \bibnamefont {{Li}}}, \bibinfo {author} {\bibfnamefont {A.~K.}\ \bibnamefont {{Meena}}}, \bibinfo {author} {\bibfnamefont {A.}~\bibnamefont {{Zitrin}}}, \bibinfo {author} {\bibfnamefont {T.~J.}\ \bibnamefont {{Broadhurst}}},\ and\ \bibinfo {author} {\bibfnamefont {A.~V.}\ \bibnamefont {{Filippenko}}},\ }\bibfield  {title} {\bibinfo {title} {{Flashlights: Properties of Highly Magnified Images Near Cluster Critical Curves in the Presence of Dark Matter Subhalos}},\ }\href {https://doi.org/10.3847/1538-4357/ad1660} {\bibfield  {journal} {\bibinfo  {journal} {\apj}\ }\textbf {\bibinfo {volume} {961}},\ \bibinfo {eid} {200}
  (\bibinfo {year} {2024})},\ \Eprint {https://arxiv.org/abs/2304.06064} {arXiv:2304.06064 [astro-ph.CO]} \BibitemShut {NoStop}%
\bibitem [{\citenamefont {{Abe}}\ \emph {et~al.}(2024)\citenamefont {{Abe}}, \citenamefont {{Kawai}},\ and\ \citenamefont {{Oguri}}}]{2024PhRvD.109h3517A}%
  \BibitemOpen
  \bibfield  {author} {\bibinfo {author} {\bibfnamefont {K.~T.}\ \bibnamefont {{Abe}}}, \bibinfo {author} {\bibfnamefont {H.}~\bibnamefont {{Kawai}}},\ and\ \bibinfo {author} {\bibfnamefont {M.}~\bibnamefont {{Oguri}}},\ }\bibfield  {title} {\bibinfo {title} {{Analytic approach to astrometric perturbations of critical curves by substructures}},\ }\href {https://doi.org/10.1103/PhysRevD.109.083517} {\bibfield  {journal} {\bibinfo  {journal} {\prd}\ }\textbf {\bibinfo {volume} {109}},\ \bibinfo {eid} {083517} (\bibinfo {year} {2024})},\ \Eprint {https://arxiv.org/abs/2311.18211} {arXiv:2311.18211 [astro-ph.CO]} \BibitemShut {NoStop}%
\bibitem [{\citenamefont {{Mao}}\ and\ \citenamefont {{Schneider}}(1998)}]{Mao1998}%
  \BibitemOpen
  \bibfield  {author} {\bibinfo {author} {\bibfnamefont {S.}~\bibnamefont {{Mao}}}\ and\ \bibinfo {author} {\bibfnamefont {P.}~\bibnamefont {{Schneider}}},\ }\bibfield  {title} {\bibinfo {title} {{Evidence for substructure in lens galaxies?}},\ }\href {https://doi.org/10.1046/j.1365-8711.1998.01319.x} {\bibfield  {journal} {\bibinfo  {journal} {Monthly Notices of the Royal Astronomical Society}\ }\textbf {\bibinfo {volume} {295}},\ \bibinfo {pages} {587} (\bibinfo {year} {1998})},\ \Eprint {https://arxiv.org/abs/astro-ph/9707187} {arXiv:astro-ph/9707187 [astro-ph]} \BibitemShut {NoStop}%
\bibitem [{\citenamefont {{Zackrisson}}\ and\ \citenamefont {{Riehm}}(2010)}]{2010AdAst2010E...9Z}%
  \BibitemOpen
  \bibfield  {author} {\bibinfo {author} {\bibfnamefont {E.}~\bibnamefont {{Zackrisson}}}\ and\ \bibinfo {author} {\bibfnamefont {T.}~\bibnamefont {{Riehm}}},\ }\bibfield  {title} {\bibinfo {title} {{Gravitational Lensing as a Probe of Cold Dark Matter Subhalos}},\ }\href {https://doi.org/10.1155/2010/478910} {\bibfield  {journal} {\bibinfo  {journal} {Advances in Astronomy}\ }\textbf {\bibinfo {volume} {2010}},\ \bibinfo {eid} {478910} (\bibinfo {year} {2010})},\ \Eprint {https://arxiv.org/abs/0905.4075} {arXiv:0905.4075 [astro-ph.CO]} \BibitemShut {NoStop}%
\bibitem [{\citenamefont {{Schechter}}\ and\ \citenamefont {{Wambsganss}}(2002)}]{Schechter2002}%
  \BibitemOpen
  \bibfield  {author} {\bibinfo {author} {\bibfnamefont {P.~L.}\ \bibnamefont {{Schechter}}}\ and\ \bibinfo {author} {\bibfnamefont {J.}~\bibnamefont {{Wambsganss}}},\ }\bibfield  {title} {\bibinfo {title} {{Quasar Microlensing at High Magnification and the Role of Dark Matter: Enhanced Fluctuations and Suppressed Saddle Points}},\ }\href {https://doi.org/10.1086/343856} {\bibfield  {journal} {\bibinfo  {journal} {The Astrophysical Journal}\ }\textbf {\bibinfo {volume} {580}},\ \bibinfo {pages} {685} (\bibinfo {year} {2002})},\ \Eprint {https://arxiv.org/abs/astro-ph/0204425} {arXiv:astro-ph/0204425 [astro-ph]} \BibitemShut {NoStop}%
\bibitem [{\citenamefont {{MacLeod}}\ \emph {et~al.}(2009)\citenamefont {{MacLeod}}, \citenamefont {{Kochanek}},\ and\ \citenamefont {{Agol}}}]{2009ApJ...699.1578M}%
  \BibitemOpen
  \bibfield  {author} {\bibinfo {author} {\bibfnamefont {C.~L.}\ \bibnamefont {{MacLeod}}}, \bibinfo {author} {\bibfnamefont {C.~S.}\ \bibnamefont {{Kochanek}}},\ and\ \bibinfo {author} {\bibfnamefont {E.}~\bibnamefont {{Agol}}},\ }\bibfield  {title} {\bibinfo {title} {{Detection of a Companion Lens Galaxy Using the Mid-Infrared Flux Ratios of the Gravitationally Lensed Quasar H1413+117}},\ }\href {https://doi.org/10.1088/0004-637X/699/2/1578} {\bibfield  {journal} {\bibinfo  {journal} {The Astrophysical Journal}\ }\textbf {\bibinfo {volume} {699}},\ \bibinfo {pages} {1578} (\bibinfo {year} {2009})},\ \Eprint {https://arxiv.org/abs/0904.0275} {arXiv:0904.0275 [astro-ph.CO]} \BibitemShut {NoStop}%
\bibitem [{\citenamefont {{Jackson}}\ \emph {et~al.}(2015)\citenamefont {{Jackson}}, \citenamefont {{Tagore}}, \citenamefont {{Roberts}}, \citenamefont {{Sluse}}, \citenamefont {{Stacey}}, \citenamefont {{Vives-Arias}}, \citenamefont {{Wucknitz}},\ and\ \citenamefont {{Volino}}}]{2015MNRAS.454..287J}%
  \BibitemOpen
  \bibfield  {author} {\bibinfo {author} {\bibfnamefont {N.}~\bibnamefont {{Jackson}}}, \bibinfo {author} {\bibfnamefont {A.~S.}\ \bibnamefont {{Tagore}}}, \bibinfo {author} {\bibfnamefont {C.}~\bibnamefont {{Roberts}}}, \bibinfo {author} {\bibfnamefont {D.}~\bibnamefont {{Sluse}}}, \bibinfo {author} {\bibfnamefont {H.}~\bibnamefont {{Stacey}}}, \bibinfo {author} {\bibfnamefont {H.}~\bibnamefont {{Vives-Arias}}}, \bibinfo {author} {\bibfnamefont {O.}~\bibnamefont {{Wucknitz}}},\ and\ \bibinfo {author} {\bibfnamefont {F.}~\bibnamefont {{Volino}}},\ }\bibfield  {title} {\bibinfo {title} {{Observations of radio-quiet quasars at 10-mas resolution by use of gravitational lensing}},\ }\href {https://doi.org/10.1093/mnras/stv1982} {\bibfield  {journal} {\bibinfo  {journal} {Monthly Notices of the Royal Astronomical Society}\ }\textbf {\bibinfo {volume} {454}},\ \bibinfo {pages} {287} (\bibinfo {year} {2015})},\ \Eprint {https://arxiv.org/abs/1508.05842} {arXiv:1508.05842 [astro-ph.GA]} \BibitemShut {NoStop}%
\bibitem [{\citenamefont {{Keeton}}\ and\ \citenamefont {{Moustakas}}(2009)}]{Keeton2009}%
  \BibitemOpen
  \bibfield  {author} {\bibinfo {author} {\bibfnamefont {C.~R.}\ \bibnamefont {{Keeton}}}\ and\ \bibinfo {author} {\bibfnamefont {L.~A.}\ \bibnamefont {{Moustakas}}},\ }\bibfield  {title} {\bibinfo {title} {{A New Channel for Detecting Dark Matter Substructure in Galaxies: Gravitational Lens Time Delays}},\ }\href {https://doi.org/10.1088/0004-637X/699/2/1720} {\bibfield  {journal} {\bibinfo  {journal} {The Astrophysical Journal}\ }\textbf {\bibinfo {volume} {699}},\ \bibinfo {pages} {1720} (\bibinfo {year} {2009})},\ \Eprint {https://arxiv.org/abs/0805.0309} {arXiv:0805.0309 [astro-ph]} \BibitemShut {NoStop}%
\bibitem [{\citenamefont {{Liu}}\ \emph {et~al.}(2024)\citenamefont {{Liu}}, \citenamefont {{Gao}}, \citenamefont {{Biesiada}},\ and\ \citenamefont {{Liao}}}]{Liu2024}%
  \BibitemOpen
  \bibfield  {author} {\bibinfo {author} {\bibfnamefont {J.}~\bibnamefont {{Liu}}}, \bibinfo {author} {\bibfnamefont {Z.}~\bibnamefont {{Gao}}}, \bibinfo {author} {\bibfnamefont {M.}~\bibnamefont {{Biesiada}}},\ and\ \bibinfo {author} {\bibfnamefont {K.}~\bibnamefont {{Liao}}},\ }\bibfield  {title} {\bibinfo {title} {{Time delay anomalies of fuzzy gravitational lenses}},\ }\href {https://doi.org/10.1103/PhysRevD.110.083536} {\bibfield  {journal} {\bibinfo  {journal} {Physical Review D}\ }\textbf {\bibinfo {volume} {110}},\ \bibinfo {eid} {083536} (\bibinfo {year} {2024})},\ \Eprint {https://arxiv.org/abs/2405.04779} {arXiv:2405.04779 [astro-ph.GA]} \BibitemShut {NoStop}%
\bibitem [{\citenamefont {{Cohn}}\ and\ \citenamefont {{Kochanek}}(2004)}]{2004ApJ...608...25C}%
  \BibitemOpen
  \bibfield  {author} {\bibinfo {author} {\bibfnamefont {J.~D.}\ \bibnamefont {{Cohn}}}\ and\ \bibinfo {author} {\bibfnamefont {C.~S.}\ \bibnamefont {{Kochanek}}},\ }\bibfield  {title} {\bibinfo {title} {{The Effects of Massive Substructures on Image Multiplicities in Gravitational Lenses}},\ }\href {https://doi.org/10.1086/392491} {\bibfield  {journal} {\bibinfo  {journal} {The Astrophysical Journal}\ }\textbf {\bibinfo {volume} {608}},\ \bibinfo {pages} {25} (\bibinfo {year} {2004})},\ \Eprint {https://arxiv.org/abs/astro-ph/0306171} {arXiv:astro-ph/0306171 [astro-ph]} \BibitemShut {NoStop}%
\bibitem [{\citenamefont {{Rigby}}\ \emph {et~al.}(2023)\citenamefont {{Rigby}} \emph {et~al.}}]{Rigby2022}%
  \BibitemOpen
  \bibfield  {author} {\bibinfo {author} {\bibfnamefont {J.}~\bibnamefont {{Rigby}}} \emph {et~al.},\ }\bibfield  {title} {\bibinfo {title} {{The Science Performance of JWST as Characterized in Commissioning}},\ }\href {https://doi.org/10.1088/1538-3873/acb293} {\bibfield  {journal} {\bibinfo  {journal} {Publications of the Astronomical Society of the Pacific}\ }\textbf {\bibinfo {volume} {135}},\ \bibinfo {eid} {048001} (\bibinfo {year} {2023})},\ \Eprint {https://arxiv.org/abs/2207.05632} {arXiv:2207.05632 [astro-ph.IM]} \BibitemShut {NoStop}%
\bibitem [{\citenamefont {{Reid}}\ and\ \citenamefont {{Honma}}(2014)}]{Reid2014}%
  \BibitemOpen
  \bibfield  {author} {\bibinfo {author} {\bibfnamefont {M.~J.}\ \bibnamefont {{Reid}}}\ and\ \bibinfo {author} {\bibfnamefont {M.}~\bibnamefont {{Honma}}},\ }\bibfield  {title} {\bibinfo {title} {{Microarcsecond Radio Astrometry}},\ }\href {https://doi.org/10.1146/annurev-astro-081913-040006} {\bibfield  {journal} {\bibinfo  {journal} {Annual Review of Astronomy and Astrophysics}\ }\textbf {\bibinfo {volume} {52}},\ \bibinfo {pages} {339} (\bibinfo {year} {2014})},\ \Eprint {https://arxiv.org/abs/1312.2871} {arXiv:1312.2871 [astro-ph.IM]} \BibitemShut {NoStop}%
\bibitem [{\citenamefont {{Liao}}\ \emph {et~al.}(2017)\citenamefont {{Liao}}, \citenamefont {{Fan}}, \citenamefont {{Ding}}, \citenamefont {{Biesiada}},\ and\ \citenamefont {{Zhu}}}]{Liao2017}%
  \BibitemOpen
  \bibfield  {author} {\bibinfo {author} {\bibfnamefont {K.}~\bibnamefont {{Liao}}}, \bibinfo {author} {\bibfnamefont {X.-L.}\ \bibnamefont {{Fan}}}, \bibinfo {author} {\bibfnamefont {X.}~\bibnamefont {{Ding}}}, \bibinfo {author} {\bibfnamefont {M.}~\bibnamefont {{Biesiada}}},\ and\ \bibinfo {author} {\bibfnamefont {Z.-H.}\ \bibnamefont {{Zhu}}},\ }\bibfield  {title} {\bibinfo {title} {{Precision cosmology from future lensed gravitational wave and electromagnetic signals}},\ }\href {https://doi.org/10.1038/s41467-017-01152-9} {\bibfield  {journal} {\bibinfo  {journal} {Nature Communications}\ }\textbf {\bibinfo {volume} {8}},\ \bibinfo {eid} {1148} (\bibinfo {year} {2017})},\ \Eprint {https://arxiv.org/abs/1703.04151} {arXiv:1703.04151 [astro-ph.CO]} \BibitemShut {NoStop}%
\bibitem [{\citenamefont {{Oguri}}(2019)}]{2019RPPh...82l6901O}%
  \BibitemOpen
  \bibfield  {author} {\bibinfo {author} {\bibfnamefont {M.}~\bibnamefont {{Oguri}}},\ }\bibfield  {title} {\bibinfo {title} {{Strong gravitational lensing of explosive transients}},\ }\href {https://doi.org/10.1088/1361-6633/ab4fc5} {\bibfield  {journal} {\bibinfo  {journal} {Reports on Progress in Physics}\ }\textbf {\bibinfo {volume} {82}},\ \bibinfo {eid} {126901} (\bibinfo {year} {2019})},\ \Eprint {https://arxiv.org/abs/1907.06830} {arXiv:1907.06830 [astro-ph.CO]} \BibitemShut {NoStop}%
\bibitem [{\citenamefont {Liao}\ \emph {et~al.}(2022)\citenamefont {Liao}, \citenamefont {Biesiada},\ and\ \citenamefont {Zhu}}]{Liao:2022}%
  \BibitemOpen
  \bibfield  {author} {\bibinfo {author} {\bibfnamefont {K.}~\bibnamefont {Liao}}, \bibinfo {author} {\bibfnamefont {M.}~\bibnamefont {Biesiada}},\ and\ \bibinfo {author} {\bibfnamefont {Z.-H.}\ \bibnamefont {Zhu}},\ }\bibfield  {title} {\bibinfo {title} {{Strongly Lensed Transient Sources: A Review}},\ }\href {https://doi.org/10.1088/0256-307X/39/11/119801} {\bibfield  {journal} {\bibinfo  {journal} {Chin. Phys. Lett.}\ }\textbf {\bibinfo {volume} {39}},\ \bibinfo {pages} {119801} (\bibinfo {year} {2022})},\ \Eprint {https://arxiv.org/abs/2207.13489} {arXiv:2207.13489 [astro-ph.HE]} \BibitemShut {NoStop}%
\bibitem [{\citenamefont {{Planck Collaboration}}\ \emph {et~al.}(2020)\citenamefont {{Planck Collaboration}} \emph {et~al.}}]{2020A&A...641A...6P}%
  \BibitemOpen
  \bibfield  {author} {\bibinfo {author} {\bibnamefont {{Planck Collaboration}}} \emph {et~al.},\ }\bibfield  {title} {\bibinfo {title} {{Planck 2018 results. VI. Cosmological parameters}},\ }\href {https://doi.org/10.1051/0004-6361/201833910} {\bibfield  {journal} {\bibinfo  {journal} {Astronomy \& Astrophysics}\ }\textbf {\bibinfo {volume} {641}},\ \bibinfo {eid} {A6} (\bibinfo {year} {2020})},\ \Eprint {https://arxiv.org/abs/1807.06209} {arXiv:1807.06209 [astro-ph.CO]} \BibitemShut {NoStop}%
\bibitem [{\citenamefont {{Navarro}}\ \emph {et~al.}(1996)\citenamefont {{Navarro}}, \citenamefont {{Frenk}},\ and\ \citenamefont {{White}}}]{1996ApJ...462..563N}%
  \BibitemOpen
  \bibfield  {author} {\bibinfo {author} {\bibfnamefont {J.~F.}\ \bibnamefont {{Navarro}}}, \bibinfo {author} {\bibfnamefont {C.~S.}\ \bibnamefont {{Frenk}}},\ and\ \bibinfo {author} {\bibfnamefont {S.~D.~M.}\ \bibnamefont {{White}}},\ }\bibfield  {title} {\bibinfo {title} {{The Structure of Cold Dark Matter Halos}},\ }\href {https://doi.org/10.1086/177173} {\bibfield  {journal} {\bibinfo  {journal} {The Astrophysical Journal}\ }\textbf {\bibinfo {volume} {462}},\ \bibinfo {pages} {563} (\bibinfo {year} {1996})},\ \Eprint {https://arxiv.org/abs/astro-ph/9508025} {arXiv:astro-ph/9508025 [astro-ph]} \BibitemShut {NoStop}%
\bibitem [{\citenamefont {{Oguri}}(2021)}]{2021PASP..133g4504O}%
  \BibitemOpen
  \bibfield  {author} {\bibinfo {author} {\bibfnamefont {M.}~\bibnamefont {{Oguri}}},\ }\bibfield  {title} {\bibinfo {title} {{Fast Calculation of Gravitational Lensing Properties of Elliptical Navarro-Frenk-White and Hernquist Density Profiles}},\ }\href {https://doi.org/10.1088/1538-3873/ac12db} {\bibfield  {journal} {\bibinfo  {journal} {Publications of the Astronomical Society of the Pacific}\ }\textbf {\bibinfo {volume} {133}},\ \bibinfo {eid} {074504} (\bibinfo {year} {2021})},\ \Eprint {https://arxiv.org/abs/2106.11464} {arXiv:2106.11464 [astro-ph.IM]} \BibitemShut {NoStop}%
\bibitem [{\citenamefont {{Hernquist}}(1990)}]{1990ApJ...356..359H}%
  \BibitemOpen
  \bibfield  {author} {\bibinfo {author} {\bibfnamefont {L.}~\bibnamefont {{Hernquist}}},\ }\bibfield  {title} {\bibinfo {title} {{An Analytical Model for Spherical Galaxies and Bulges}},\ }\href {https://doi.org/10.1086/168845} {\bibfield  {journal} {\bibinfo  {journal} {The Astrophysical Journal}\ }\textbf {\bibinfo {volume} {356}},\ \bibinfo {pages} {359} (\bibinfo {year} {1990})}\BibitemShut {NoStop}%
\bibitem [{\citenamefont {{Birrer}}\ and\ \citenamefont {{Amara}}(2018)}]{2018ascl.soft04012B}%
  \BibitemOpen
  \bibfield  {author} {\bibinfo {author} {\bibfnamefont {S.}~\bibnamefont {{Birrer}}}\ and\ \bibinfo {author} {\bibfnamefont {A.}~\bibnamefont {{Amara}}},\ }\href@noop {} {\bibinfo {title} {{Lenstronomy: Multi-purpose gravitational lens modeling software package}}},\ \bibinfo {howpublished} {Astrophysics Source Code Library, record ascl:1804.012} (\bibinfo {year} {2018}),\ \Eprint {https://arxiv.org/abs/1804.012} {ascl:1804.012} \BibitemShut {NoStop}%
\bibitem [{\citenamefont {{Birrer}}\ \emph {et~al.}(2021)\citenamefont {{Birrer}} \emph {et~al.}}]{2021JOSS....6.3283B}%
  \BibitemOpen
  \bibfield  {author} {\bibinfo {author} {\bibfnamefont {S.}~\bibnamefont {{Birrer}}} \emph {et~al.},\ }\bibfield  {title} {\bibinfo {title} {{lenstronomy II: A gravitational lensing software ecosystem}},\ }\href {https://doi.org/10.21105/joss.03283} {\bibfield  {journal} {\bibinfo  {journal} {The Journal of Open Source Software}\ }\textbf {\bibinfo {volume} {6}},\ \bibinfo {eid} {3283} (\bibinfo {year} {2021})},\ \Eprint {https://arxiv.org/abs/2106.05976} {arXiv:2106.05976 [astro-ph.CO]} \BibitemShut {NoStop}%
\bibitem [{\citenamefont {{Schive}}\ \emph {et~al.}(2014)\citenamefont {{Schive}}, \citenamefont {{Liao}}, \citenamefont {{Woo}}, \citenamefont {{Wong}}, \citenamefont {{Chiueh}}, \citenamefont {{Broadhurst}},\ and\ \citenamefont {{Hwang}}}]{2014PhRvL.113z1302S}%
  \BibitemOpen
  \bibfield  {author} {\bibinfo {author} {\bibfnamefont {H.-Y.}\ \bibnamefont {{Schive}}}, \bibinfo {author} {\bibfnamefont {M.-H.}\ \bibnamefont {{Liao}}}, \bibinfo {author} {\bibfnamefont {T.-P.}\ \bibnamefont {{Woo}}}, \bibinfo {author} {\bibfnamefont {S.-K.}\ \bibnamefont {{Wong}}}, \bibinfo {author} {\bibfnamefont {T.}~\bibnamefont {{Chiueh}}}, \bibinfo {author} {\bibfnamefont {T.}~\bibnamefont {{Broadhurst}}},\ and\ \bibinfo {author} {\bibfnamefont {W.~Y.~P.}\ \bibnamefont {{Hwang}}},\ }\bibfield  {title} {\bibinfo {title} {{Understanding the Core-Halo Relation of Quantum Wave Dark Matter from 3D Simulations}},\ }\href {https://doi.org/10.1103/PhysRevLett.113.261302} {\bibfield  {journal} {\bibinfo  {journal} {Physical Review Letters}\ }\textbf {\bibinfo {volume} {113}},\ \bibinfo {eid} {261302} (\bibinfo {year} {2014})},\ \Eprint {https://arxiv.org/abs/1407.7762} {arXiv:1407.7762 [astro-ph.GA]} \BibitemShut {NoStop}%
\bibitem [{\citenamefont {{Schive}}\ \emph {et~al.}(2016)\citenamefont {{Schive}}, \citenamefont {{Chiueh}}, \citenamefont {{Broadhurst}},\ and\ \citenamefont {{Huang}}}]{2016ApJ...818...89S}%
  \BibitemOpen
  \bibfield  {author} {\bibinfo {author} {\bibfnamefont {H.-Y.}\ \bibnamefont {{Schive}}}, \bibinfo {author} {\bibfnamefont {T.}~\bibnamefont {{Chiueh}}}, \bibinfo {author} {\bibfnamefont {T.}~\bibnamefont {{Broadhurst}}},\ and\ \bibinfo {author} {\bibfnamefont {K.-W.}\ \bibnamefont {{Huang}}},\ }\bibfield  {title} {\bibinfo {title} {{Contrasting Galaxy Formation from Quantum Wave Dark Matter, {\ensuremath{\psi}}DM, with {\ensuremath{\Lambda}}CDM, using Planck and Hubble Data}},\ }\href {https://doi.org/10.3847/0004-637X/818/1/89} {\bibfield  {journal} {\bibinfo  {journal} {The Astrophysical Journal}\ }\textbf {\bibinfo {volume} {818}},\ \bibinfo {eid} {89} (\bibinfo {year} {2016})},\ \Eprint {https://arxiv.org/abs/1508.04621} {arXiv:1508.04621 [astro-ph.GA]} \BibitemShut {NoStop}%
\bibitem [{\citenamefont {{Chan}}\ \emph {et~al.}(2020)\citenamefont {{Chan}}, \citenamefont {{Schive}}, \citenamefont {{Wong}}, \citenamefont {{Chiueh}},\ and\ \citenamefont {{Broadhurst}}}]{2020PhRvL.125k1102C}%
  \BibitemOpen
  \bibfield  {author} {\bibinfo {author} {\bibfnamefont {J.~H.~H.}\ \bibnamefont {{Chan}}}, \bibinfo {author} {\bibfnamefont {H.-Y.}\ \bibnamefont {{Schive}}}, \bibinfo {author} {\bibfnamefont {S.-K.}\ \bibnamefont {{Wong}}}, \bibinfo {author} {\bibfnamefont {T.}~\bibnamefont {{Chiueh}}},\ and\ \bibinfo {author} {\bibfnamefont {T.}~\bibnamefont {{Broadhurst}}},\ }\bibfield  {title} {\bibinfo {title} {{Multiple Images and Flux Ratio Anomaly of Fuzzy Gravitational Lenses}},\ }\href {https://doi.org/10.1103/PhysRevLett.125.111102} {\bibfield  {journal} {\bibinfo  {journal} {Physical Review Letters}\ }\textbf {\bibinfo {volume} {125}},\ \bibinfo {eid} {111102} (\bibinfo {year} {2020})},\ \Eprint {https://arxiv.org/abs/2002.10473} {arXiv:2002.10473 [astro-ph.GA]} \BibitemShut {NoStop}%
\bibitem [{\citenamefont {{Abe}}\ \emph {et~al.}(2025)\citenamefont {{Abe}}, \citenamefont {{Oguri}}, \citenamefont {{Birrer}}, \citenamefont {{Khadka}}, \citenamefont {{Marshall}}, \citenamefont {{Lemon}}, \citenamefont {{More}},\ and\ \citenamefont {{LSST Dark Energy Science Collaboration}}}]{2025OJAp....8E...8A}%
  \BibitemOpen
  \bibfield  {author} {\bibinfo {author} {\bibfnamefont {K.~T.}\ \bibnamefont {{Abe}}}, \bibinfo {author} {\bibfnamefont {M.}~\bibnamefont {{Oguri}}}, \bibinfo {author} {\bibfnamefont {S.}~\bibnamefont {{Birrer}}}, \bibinfo {author} {\bibfnamefont {N.}~\bibnamefont {{Khadka}}}, \bibinfo {author} {\bibfnamefont {P.~J.}\ \bibnamefont {{Marshall}}}, \bibinfo {author} {\bibfnamefont {C.}~\bibnamefont {{Lemon}}}, \bibinfo {author} {\bibfnamefont {A.}~\bibnamefont {{More}}},\ and\ \bibinfo {author} {\bibnamefont {{LSST Dark Energy Science Collaboration}}},\ }\bibfield  {title} {\bibinfo {title} {{A halo model approach for mock catalogs of time-variable strong gravitational lenses}},\ }\href {https://doi.org/10.33232/001c.128482} {\bibfield  {journal} {\bibinfo  {journal} {The Open Journal of Astrophysics}\ }\textbf {\bibinfo {volume} {8}},\ \bibinfo {eid} {8} (\bibinfo {year} {2025})},\ \Eprint {https://arxiv.org/abs/2411.07509} {arXiv:2411.07509 [astro-ph.CO]} \BibitemShut {NoStop}%
\bibitem [{\citenamefont {{Ivezi{\'c}}}\ \emph {et~al.}(2019)\citenamefont {{Ivezi{\'c}}} \emph {et~al.}}]{2019ApJ...873..111I}%
  \BibitemOpen
  \bibfield  {author} {\bibinfo {author} {\bibnamefont {{Ivezi{\'c}}}} \emph {et~al.},\ }\bibfield  {title} {\bibinfo {title} {{LSST: From Science Drivers to Reference Design and Anticipated Data Products}},\ }\href {https://doi.org/10.3847/1538-4357/ab042c} {\bibfield  {journal} {\bibinfo  {journal} {The Astrophysical Journal}\ }\textbf {\bibinfo {volume} {873}},\ \bibinfo {eid} {111} (\bibinfo {year} {2019})},\ \Eprint {https://arxiv.org/abs/0805.2366} {arXiv:0805.2366 [astro-ph]} \BibitemShut {NoStop}%
\bibitem [{\citenamefont {{Foreman-Mackey}}\ \emph {et~al.}(2013)\citenamefont {{Foreman-Mackey}}, \citenamefont {{Conley}}, \citenamefont {{Meierjurgen Farr}}, \citenamefont {{Hogg}}, \citenamefont {{Lang}}, \citenamefont {{Marshall}}, \citenamefont {{Price-Whelan}}, \citenamefont {{Sanders}},\ and\ \citenamefont {{Zuntz}}}]{2013ascl.soft03002F}%
  \BibitemOpen
  \bibfield  {author} {\bibinfo {author} {\bibfnamefont {D.}~\bibnamefont {{Foreman-Mackey}}}, \bibinfo {author} {\bibfnamefont {A.}~\bibnamefont {{Conley}}}, \bibinfo {author} {\bibfnamefont {W.}~\bibnamefont {{Meierjurgen Farr}}}, \bibinfo {author} {\bibfnamefont {D.~W.}\ \bibnamefont {{Hogg}}}, \bibinfo {author} {\bibfnamefont {D.}~\bibnamefont {{Lang}}}, \bibinfo {author} {\bibfnamefont {P.}~\bibnamefont {{Marshall}}}, \bibinfo {author} {\bibfnamefont {A.}~\bibnamefont {{Price-Whelan}}}, \bibinfo {author} {\bibfnamefont {J.}~\bibnamefont {{Sanders}}},\ and\ \bibinfo {author} {\bibfnamefont {J.}~\bibnamefont {{Zuntz}}},\ }\href@noop {} {\bibinfo {title} {{emcee: The MCMC Hammer}}},\ \bibinfo {howpublished} {Astrophysics Source Code Library, record ascl:1303.002} (\bibinfo {year} {2013}),\ \Eprint {https://arxiv.org/abs/1303.002} {ascl:1303.002} \BibitemShut {NoStop}%
\end{thebibliography}%

\end{document}